\documentclass[12pt]{article}

\usepackage{scicite}
\usepackage{times}

\usepackage{float}
\usepackage{graphicx}
\graphicspath{{Figures/}}
\usepackage{caption}

\newenvironment{sciabstract}{%
	\begin{quote} \bf}
	{\end{quote}}

\title{Universality of macroscopic neuronal dynamics in $\textit{Caenorhabditis elegans}$.}

\author{Connor Brennan$^1$ \& Alex Proekt$^2$\\
	\\
	\normalsize{$^{1}$Department of Neuroscience, University of Pennsylvania}\\
	\normalsize{$^{2}$Department of Anesthesiology and Critical Care, University of Pennsylvania}\\
	\\
	\normalsize{$^\ast$To whom correspondence should be addressed; E-mail:  brenco@mail.med.upenn.edu.}
}

\date{}

\begin{document}	
	\baselineskip24pt
	
	
	\maketitle 
	
	\begin{sciabstract}
		Recordings of whole brain activity with single neuron resolution are now feasible in simple organisms. Yet, it is still challenging to appropriately simplify such complex, noisy, and multivariate data in order to reveal general principles of nervous system function. Here, we develop a method that allows us to extract global brain dynamics from pan-neuronal imaging. Success of this method is rooted in a surprising mathematical connection between dimensionality reduction and a general class of thermodynamic systems. Application of this theoretical framework to the nervous system of \textit{C. elegans} reveals the manifold that sculpts global brain dynamics. This manifold allows us to predict switches between worm behaviors across individuals, implying that macroscopic dynamics embodied by the manifold are universal. In contrast, activation of individual neurons differs consistently between worms. These findings suggest that brains of genetically identical individuals express distinct microscopic neuronal configurations which nonetheless yield equivalent macroscopic dynamics. 
	\end{sciabstract}

	Both deterministic\cite{hodgkin1952currents,hodgkin1952measurement} and stochastic\cite{schneidman2006weak,schneidman2003network,tkavcik2013simplest} models have been successfully used to understand brain function.  Detailed biophysical models can be used to faithfully model voltage dynamics in single neurons\cite{hodgkin1952currents,hodgkin1952measurement, fitzhugh1960thresholds, golowasch2002failure} while  pairwise correlations between firing of individual neurons can predict the likelihood of  global firing patterns\cite{schneidman2006weak,schneidman2003network,tkavcik2013simplest}. Yet, detailed biophysical models of even simple brains are experimentally intractable\cite{gjorgjieva2016computational}, computationally costly\cite{izhikevich2003simple,markram2015reconstruction}, and not necessarily conceptually revealing\cite{gjorgjieva2016computational,selverston1980central,barral2016synaptic}. On the other hand, stochastic models are not sufficient to reproduce the observed macroscopic dynamics \cite{tang2008maximum,ohiorhenuan2010sparse}. Significant simplification is necessary in order to arrive at a model of the nervous system at the behaviorally-relevant macroscopic level. Yet, it is not clear how the variability  observed at the microscopic level in the nervous system can be appropriately simplified. 
	
	Collective macroscopic properties are commonly understood as emergent epiphenomena that arise as a consequence of microscopic interactions but do not themselves affect the system\cite{campbell1974downward}. Yet, from an evolutionary standpoint, selection operates on the level of organismal behavior\cite{lassig2008biological} mediated by macroscopic brain dynamics rather than individual ion channels or neurons. Thus, here we hypothesized that this evolutionary constraint will manifest as invariance of macroscopic brain dynamics. However, because multiple microscopic configurations can give rise to equivalent macroscopic dynamics it is possible for individual components of the system to vary across individuals.

	We test this hypothesis in nematode \textit{C. elegans} because of its unique advantages as a model organism. All 302 neurons\cite{white1986structure} and all of their connections are known\cite{izquierdo2013connecting,bargmann2013connectome,varshney2011structural}. Simultaneous recordings of the majority of the neurons in the brain (head ganglia) of the worm have been performed in vivo\cite{kato2015global,nguyen2016whole,prevedel2014simultaneous,tian2009imaging} (Figure~\ref{Fig:1}A, Supplement S1) and made publicly available. Motor commands of the worm can be inferred from activation of well-characterized neurons\cite{kato2015global,li2014encoding,luo2014dynamic,larsch2013high}. For instance activation of the AVA neuron can be used to infer the direction of locomotion (Figure~\ref{Fig:1}B).


	Changes in behavior are associated with coherent changes in the activity of multiple neurons\cite{michaels2016neural,fetz1992movement,georgopoulos1986neuronal,salinas2001correlated} even in the simple nervous system of \textit{C. elegans}\cite{kato2015global} (Figure~\ref{Fig:1}). Correlations among neurons can be used to reduce the dimensionality of neuronal activity using principal component analysis (PCA)\cite{kato2015global} (Figures \ref{Fig:pcaProjections} and \ref{Fig:reconstruction}). Indeed, in the nervous system of \textit{C. elegans} only two principal components (PCs) are required to extract approximately 60\% of the fluctuations in neuronal activity\cite{kato2015global}. 
	
	There is, however, a fundamental distinction between neuronal activity and neuronal dynamics\cite{churchland2012neural,salinas2001correlated}. Neuronal activity is an output of the dynamical system governed by the biophysics of individual neurons and their connections\cite{seung1996brain,beer1995dynamical,miller1982mechanisms}. These biophysical processes influence neuronal activity and are in turn influenced by it. Yet they are not directly apparent in recordings of neuronal activity.
	
	Neuronal dynamics  can be thought of as a trajectory through the space spanned by the relevant variables. This trajectory can never cross itself in a deterministic dynamical system\cite{strogatz2014nonlinear}. In real systems, however, measurement noise and stochastic processes lead to some tangling of the trajectories. Thus, a combination of deterministic and stochastic approaches is necessary. Here we successfully combine deterministic and stochastic approaches to discover trajectory bundles\cite{sugihara2012detecting} that together form a manifold. This manifold governs macroscopic dynamics of the \textit{C. elegans} nervous system. Our key finding is that the shape of this manifold is conserved among individuals. This universality allows us to make predictions concerning future behaviors of the worm. Remarkably these predictions are valid across individuals despite the fact that activations of some identified neurons are consistently and qualitatively different among worms within a clonal population. 
	
		\begin{figure}[H]
			\centering
			\includegraphics[width=\textwidth]{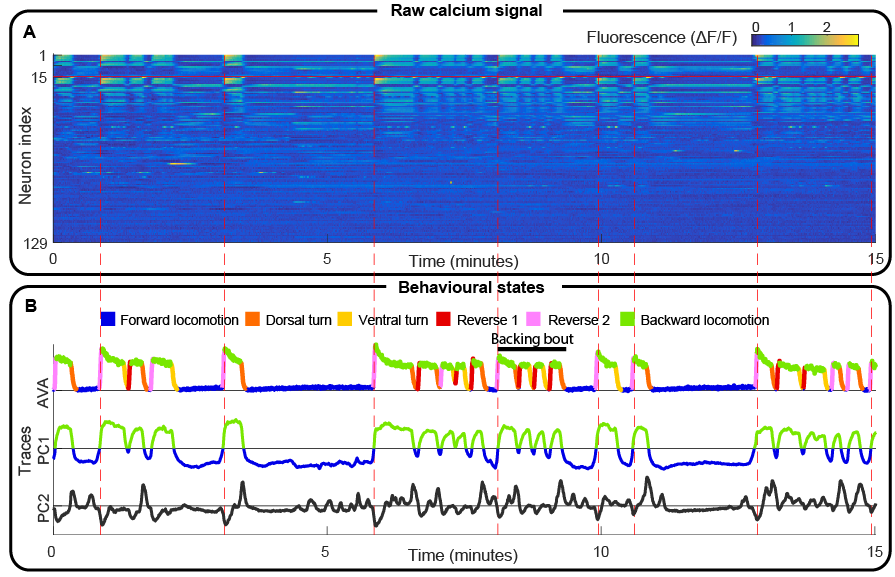}
			\phantomcaption
		\end{figure}
		\begin{figure}[H]
			\ContinuedFloat
			\caption{\textbf{behavioral assignment of \textit{C. elegans} pan-neuronal imaging}
				\small{\textbf{A)} Heat plot of fluorescence ($\Delta$F/F) for 129 neurons in \textit{C. elegans} head ganglia recorded by Kato et al\cite{kato2015global}. Neurons 1 to 15 (above horizontal red line) were identified in all five individual worms\cite{kato2015global}. For the purposes of comparing activity between worms this measure was normalized and expressed as z-score. \textbf{B)} Trace of AVA neuron colored according to the inferred behavioral state (top)\cite{kato2015global}. Solid black line shows the duration of a sample backing bout, defined as repeated backing behavior uninterrupted by forward locomotion. Vertical red lines show beginning of a backing bout. Neuronal activity in \textbf{A} projected onto the first two principal components (PC1 and PC2) (bottom). Turns and reversals are not easily assigned on the basis of PC1 alone. Thus, only forward and backward locomotion are labeled (blue and green respectively).}
				\label{Fig:1}}
		\end{figure}
		
		\clearpage
	
	\section*{Stochastic processes alone are insufficient to model single worm dynamics.}
	
	Assuming only that the processes giving rise to the observed neuronal activity do not change appreciably during the time course of the experiment, and given the constraints of observed neuronal statistics\cite{schneidman2003network,schneidman2006weak}, the simplest  model of neuronal dynamics is Brownian motion (Supplement S2)\cite{pathria1996statistical}.  To construct this model, we estimate the probability density function $P$ (Figure~\ref{Fig:2}A, middle panel), by binning neuronal activity (Figure~\ref{Fig:2}A, bottom) in the plane spanned by the first two principal components (PC1 and PC2). Brownian motion given by 
	\begin{equation} \label{Brownian}
	\frac{d \mathbf{X}}{d t} = D\frac{\nabla P(\mathbf{X})}{P(\mathbf{X})} + \epsilon,		    
	\end{equation}
	is determined entirely by $P$. Diffusion constant $D$ and noise $\epsilon$ set the temporal scale but do not influence the dynamics. In this case $\mathbf{X}$ is the position vector in the plane spanned by PC1 and PC2.  Dashed arrows (Figure~\ref{Fig:2}A top panel) show the flux of neuronal activity predicted by Brownian motion -- preferred direction always tends towards the local energy minimum.  While  simulations of Brownian motion recapitulate the statistics, they fail to qualitatively or quantitatively recapitulate the observed neuronal or behavioral dynamics even for a single worm (Figures \ref{Fig:2}B and \ref{Fig:2}C, respectively). Thus, stochastic processes alone are unable to give rise to the observed neuronal dynamics. 
	
	\section*{Local flux of neuronal activity is sufficient to model a single worm but fails to generalize across worms.}
	
	The flux of neuronal activity computed as the observed transition probability between neighboring bins rather than the gradient of $P$ (Eq.~\ref{Brownian}) is shown in  Figure~\ref{Fig:2}D. We refer to the dynamics given by these experimentally observed transition probabilities  as the Empirical Markov Model (EMM). EMM simulations not only recapitulate the observed statistics (Figure~\ref{Fig:2}D, top) but also the qualitative and quantitative aspects of neuronal dynamics (Figure~\ref{Fig:2}E-F) within a single worm. Near energy minima both models have similar fluxes (red arrows, Figures \ref{Fig:2}A and \ref{Fig:2}D). EMM is distinguished, however, by the presence of an additional flux component (blue arrows, Figure~\ref{Fig:2}D) (Figure~\ref{Fig:quotient}). Because this additional component cannot be inferred from $P$ alone, we refer to it as deterministic. Note that the deterministic component of the flux is cyclic (Figures~\ref{Fig:2}D and \ref{Fig:quotient}).  
	
	Surprisingly, attempts to model neuronal dynamics by directly projecting all worms onto a common PCA-based coordinate system  followed by application of EMM  fail catastrophically (Figure~\ref{Fig:2}G) despite being successful in each worm individually (Figure~\ref{Fig:EMMStats}). One reason for this failure is that only 15 neurons were consistently identified across worms (Figure~\ref{Fig:EMMStats}). The primary problem, however, is that projecting all worms onto a common coordinate system leads to tangling of trajectories (Figure~\ref{Fig:allPca} and \ref{Fig:allPcaDiff}). Consequently, the flux computed from the cross-worm EMM model is similar to that predicted by Brownian motion (Figure~\ref{Fig:quoitent15} and \ref{Fig:quoitent15Diff}). While including more principal components could in principle improve generalizability of EMM, the amount of data needed to constrain $P$ grows exponentially with the number of dimensions and quickly becomes experimentally intractable\cite{aggarwal2001surprising}. Thus, in order to be experimentally tractable, dynamics have to lie in a low dimensional manifold. Yet, lack of generalizability of EMM illustrates that there is no simple linear relationship between neuronal activity and the macroscopic variables that sculpt neuronal dynamics.    
	
	\begin{figure}[H]
		\centering
		\includegraphics[width=\textwidth]{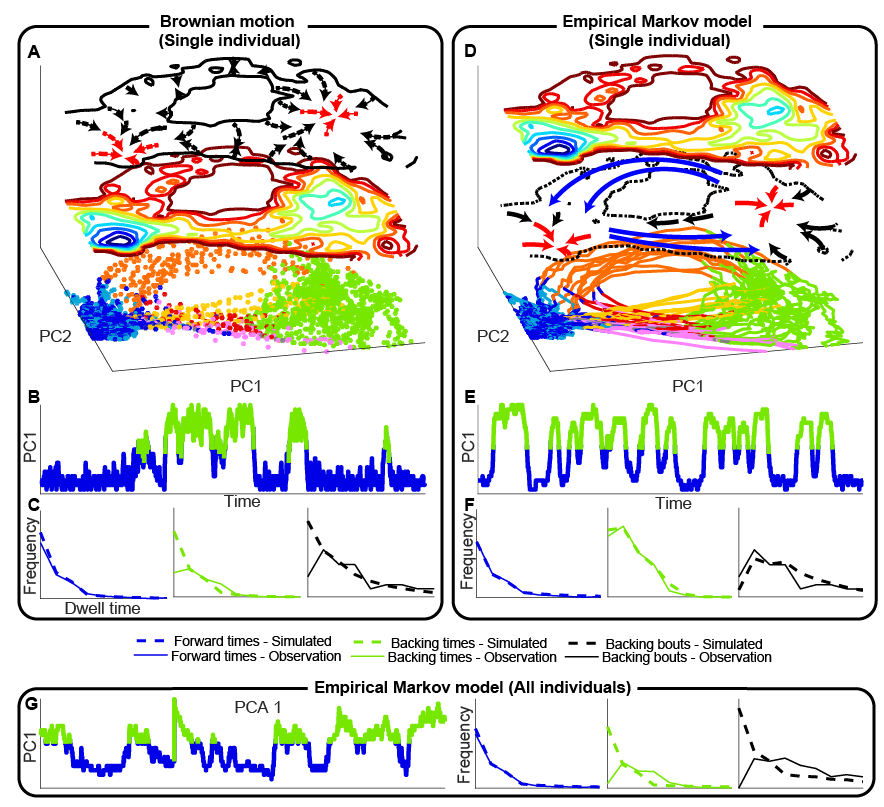}
		\phantomcaption
	\end{figure}
	\begin{figure}[H]
		\ContinuedFloat
		\caption{\textbf{Neuronal dynamics as diffusion in neuronal activity space}
			\small{\textbf{A-C)} Brownian motion is not sufficient to capture neuronal dynamics in a single worm. \textbf{A)} Neuronal activity projected onto PC1 and 2 and colored according to the behavioral state (bottom). Empirically derived distribution of points (middle). Blue shows high probability (low energy). Red shows low probability (high energy). Brownian motion predicts that neuronal activity will flow down the gradient of the distribution (arrows top). \textbf{B)} Simulated neuronal activity (only PC1 shown) is qualitatively different from observed neuronal dynamics (Figure~\ref{Fig:1}B) \textbf{C)} Simulated dwell times in different behavioral states differ from those observed experimentally. \textbf{D-F)} Empirical Markov model (EMM) captures dynamics in a single worm. \textbf{D)} Neuronal activity projected as in \textbf{A-C}. Unlike in \textbf{A-C}, individual snapshots are now connected according to the order in which they were empirically observed. Empirically observed transition probabilities are shown as flow (middle). The distribution computed by simulating empirically observed transition probabilities (\textbf{D}, top) is identical to the empirically observed distribution (\textbf{A}, middle). \textbf{E)} Simulation of neuronal activity projected onto PC1 qualitatively recapitulates observed dynamics. \textbf{F)} Dwell time distributions of behavioral states match those observed empirically. \textbf{G)} Empirical Markov model fails to generalize across worms. Projection of recordings from the 15 neurons identified across all worms onto common PCs fails to give rise to a meaningful EMM. Simulated PC1 (left) and predicted dwell time distributions of behavioral states (right) are qualitatively and quantitatively different from the observed neuronal activity.}
			\label{Fig:2}}
	\end{figure}
	
	\clearpage
	
	\section*{Macroscopic dynamics are conserved among worms.}
	
	While worms may differ in terms of specific microscopic details of neuronal activation, we hypothesized that at the collective macroscopic level dynamics will be conserved between worms. This hypothesis asserts that when projected onto a suitable coordinate system the trajectories of neuronal activity, \textit{i.e.} the manifolds, have similar shapes in different individuals (Figure~\ref{Fig:3}). This coordinate system, known as the phase space, contains all of the relevant dynamical variables. 
	
	Consistent with the hypothesis, we identified a phase space in which the manifold constructed by projecting activity of 107 neurons in a single worm  (Figure~\ref{Fig:3}A) is essentially identical to the manifold constructed across individual worms using just the common set of 15 neurons (Figure~\ref{Fig:3}B) (Supplement S4). To construct the manifolds in Figure~\ref{Fig:3} we averaged neuronal trajectory with respect to the phase of the cyclic flux rather than with respect to neuronal activity (Figures~ \ref{Fig:2}G,~\ref{Fig:allPca}-\ref{Fig:quoitent15Diff},). Thus, phase of the cyclic flux (arrow) is the first relevant macroscopic dynamical variable. In addition to the phase, the only other relevant variable is  identity of the flux -- specifying which loop the system is currently in. Therefore, the phase space of the \textit{C. elegans} nervous system is only two dimensional (Figure \ref{Fig:manifoldSpace}). 
	
	The separation of trajectories in phase space allows for near perfect prediction of behavioral statistics (Figure \ref{Fig:3}C-D)  as well as efficient simulations of whole-brain dynamics (Figure \ref{Fig:reconstruction}) valid for each worm individually and collectively across worms. To further strengthen the argument for manifold invariance across worms, we constructed a manifold based on the data from four out of the five worms and used it to predict the behavior of the fifth worm not used in the construction of the manifold (Figure~\ref{Fig:3}E, \ref{Fig:mainifoldStats}, \ref{Fig:leftOutStats}).


	Remarkably, the manifold provides meaningful information not just about statistics of behaviors but also about behavioral transitions on a cycle-by-cycle basis. In  (Figures \ref{Fig:3}F-I) we focus on predicting the termination of backwards locomotion (transition from green to yellow/orange). Figures~\ref{Fig:forwardPhase} and \ref{Fig:forwardEntropy} show other behavioral transitions. In Figure~\ref{Fig:3}G, we compute the time until termination as a function of phase along the manifold (Supplement S3). As expected, the time to behavioral transition decreases as the phase evolves towards the transition zone between behaviors (orange trace). This cannot be predicted based on dwell time statistics alone (blue trace). As the system nears the point of transition, the influence of the stochasticity decreases and uncertainty (entropy) of the prediction decreases concomitantly (Figure~\ref{Fig:3}H orange). This information is also not contained in dwell time statistics (Figures~\ref{Fig:3}H blue trace,  \ref{Fig:forwardPhase} and  \ref{Fig:forwardEntropy}). 
	
	There is a key distinction between neuronal activity and neuronal dynamics. Neuronal activity changes abruptly (Figure \ref{Fig:1}) and transitions appear stochastic. In contrast,  the manifold space is smooth. This is akin to smooth Hodgkin Huxley variables giving rise to abrupt changes in membrane voltage such as action and plateau potentials\cite{fitzhugh1960thresholds}. The smoothness of the manifold exposes the deterministic dynamics that depend predominantly on the phase of the cyclic flux. Yet this phase is not readily apparent in observations of neuronal activity.

	
	\begin{figure}[H]
		\centering
		\includegraphics[width=\textwidth]{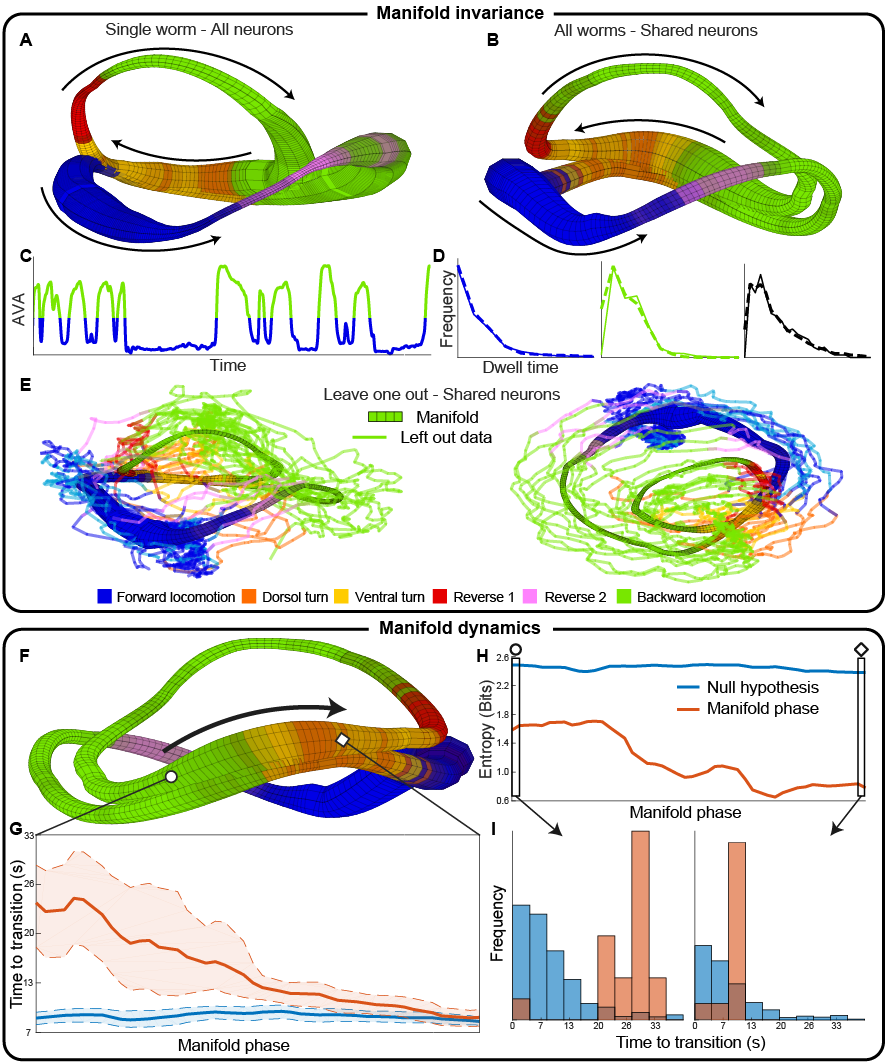}
		\phantomcaption
	\end{figure}
	\begin{figure}[H]
		\ContinuedFloat
		\caption{\textbf{Universality of manifold allows cross worm predictions.}
			\small{\textbf{A)} Manifold of coherent trajectory bundles for a single worm computed on the basis of 100+ neurons. Color shows behavioral state, width shows point density, arrows show the preferred direction of motion along each loop. Neuronal trajectory averaged with respect to phase of the flux is projected onto the first 3 PCs. PCA is not required for the simulations or quantitative predictions and is solely used for illustration \textbf{B)} Manifold of neuronal dynamics computed for all worms using only the 15 neurons shared across worms. The shape of the manifold in \textbf{B} is essentially identical to \textbf{A}. \textbf{C)} AVA neuron simulated using the all-worm manifold (Figure~\ref{Fig:3}B) is qualitatively similar to observed AVA activity (Figure~\ref{Fig:1}B). \textbf{D)} Statistics of simulated behavioral dwell times are essentially identical to observed behavioral statistics (forward blue; backward green; backing bouts black).   \textbf{E)} Manifold constructed on the basis of 4 worms. Raw data from the left out fifth worm is superimposed onto the manifold. Manifold in the same orientation as \textbf{A} and \textbf{B} (left). Manifold rotated to highlight the coherence in trajectories (right). \textbf{F-I)} Predictions of behavioral transitions on a cycle-by-cycle basis. \textbf{F)} Evolution of the system from the circle to the diamond marker is tracked by phase along the cyclic flux. \textbf{G)} Each point on the manifold is identified by the time since the initiation of backward locomotion and independently by the phase of the cyclic flux. The former gives rise to a prediction for termination of backwards locomotion based on behavioral statistics alone (mean shown by blue line, shading shows 95\% confidence interval). The latter gives rise to the prediction based on the manifold  (mean shown by orange line, shading shows 95\% confidence interval). Time to termination is progressively shortened as the phase proceeds from the circle to the diamond marker. This information is not contained in the behavioral statistics alone. \textbf{H)} Uncertainty (entropy) about expected time until termination of backward locomotion for manifold-based model (orange) is lower than for dwell time-based model (blue) for all phases of the manifold. Uncertainty decreases progressively as phase evolves from circle to diamond. \textbf{I)} Empirical distributions of time until termination at the start (circle) and end (diamond) of the subsection.}
			\label{Fig:3}}
	\end{figure}
	
	\clearpage
	
	\section*{Construction of an asymmetrical transition probability matrix to identify the phase of the flux of neuronal activity.}
	
	Altogether, findings in Figure~\ref{Fig:3} strongly argue that the phase of the cyclic flux is a key macroscopic dynamical variable of \textit{C. elegans} nervous system. Cyclicity of neuronal dynamics (Figures \ref{Fig:3}, \ref{Fig:2}D and \ref{Fig:quotient}) is not accidental  -- it naturally arises in stochastic differential equations. To show this, we start with the law of probability conservation
	\begin{equation} \label{Conservation}
	\frac{dP(\mathbf{X})}{dt} = -\nabla \mathbf{J} (\mathbf{X},t).
	\end{equation}
	Where $\mathbf{J}(\mathbf{X},t)$ is the flux at state $\mathbf{X}$ and time $t$. We first assume steady state $\frac{dP(\mathbf{X})}{dt}=0$ and then use the Fokker-Planck equation\cite{pathria1996statistical} -- which describes the time-evolution of driven stochastic systems -- to arrive (Supplement S2) at the stochastic partial differential equation that governs the temporal evolution of the broad class of systems with both stochastic and deterministic processes 
	\begin{equation} \label{EMM}
	\frac{d \mathbf{X}}{d t} = D\frac{\nabla P(\mathbf{X})}{P(\mathbf{X})} + \frac{\mathbf{J}(\mathbf{X})}{P(\mathbf{X})} + \epsilon.
	\end{equation}
	The first term in Eq.~\ref{EMM} is Brownian motion. The second term contains the deterministic flux $\mathbf{J}$. Brownian motion is the trivial solution  ($\mathbf{J} = 0$) often assumed in stochastic models\cite{schneidman2006weak}. The non-trivial solution ($\mathbf{J} \neq 0$) satisfies the steady state assumption so long as the flux is cyclic  -- that is, if the net flux along any loop is zero\cite{wang2008potential,yan2013nonequilibrium}. Thus, divergence-free cyclic flux is the most parsimonious solution that at once satisfies the steady state assumption and allows for deterministic dynamics. 
	
	Since Eq.~\ref{EMM} naturally arises in reaction-diffusion systems, we adapt diffusion mapping\cite{nadler2006diffusion,coifman2006diffusion,lian2015multivariate} to express distances between two different states of the nervous system ($\mathbf{X}$ and $\mathbf{Y}$) as the probability of transition between them 
	\begin{equation} \label{dmkernel}
	P_{X\rightarrow Y}=exp\left (-\frac{\left \| \mathbf{X} - \mathbf{Y} \right \|_2^2}{4\varepsilon }  \right ),
	\end{equation} 
	where $\varepsilon$ is a measure of the size of the local neighborhood and $\left \| \cdot \right \|_2$ is the Euclidean distance. Eq.~\ref{dmkernel} describes a Gaussian distribution which peaks at $\mathbf{X}$ and decays as a function of distance from it. Transition probabilities between all pairs of states can be summarized in a transition probability matrix $\mathbf{M}$ commonly referred to as a diffusion map (DM)  because of the analogy between diffusion and transition probability (Supplement S4). 
	
	Yet, by construction $\mathbf{M}$ is symmetric $P_{X\rightarrow Y} = P_{Y\rightarrow X}$ and therefore compatible only with Brownian motion ($\mathbf{J} = 0$).  This symmetry arises because Eq.~\ref{dmkernel} does not take into account the order in which states $\mathbf{X}$ and $\mathbf{Y}$ are observed. To illustrate how DMs can be modified to include dynamics we  use a simple system (Supplement S5) governed by Eq.~\ref{EMM} (Figure~\ref{Fig:4}A).    
	
	We denote the state of the system at time $t$ as $\mathbf{X}_t$. Because of the cyclic flux, the system can readily transition from $\mathbf{X}_t$ to $\mathbf{X}_{t+1}$ but not backwards (Figure~\ref{Fig:4}C). We use this simple intuition to modify DMs as 
	\begin{equation} \label{DiffusionMap}
	P_{\mathbf{X}_t\rightarrow \mathbf{Y}} =exp\left (-\frac{\left \| \mathbf{X}_{t+1}-\mathbf{Y} \right \|_2^2}{2 \sigma^2}  \right ).
	\end{equation}
	The key difference from Eq.~\ref{dmkernel} is that the transition probability peaks at the next empirically observed state of the system (Figure~\ref{Fig:4}C). $\sigma$ in Eq~\ref{DiffusionMap} is a data-based estimate of the size of the local neighborhood (Supplement S4). Thus, the system in state $\mathbf{X}_{t}$ will most likely transition to the next empirically observed state $\mathbf{X}_{t+1}$, but may alternatively transition to a state near $\mathbf{X}_{t+1}$, corresponding to a different nearby trajectory (Figure~\ref{Fig:4}D). The transition probability after $N$ steps, given by $\mathbf{M}^N$ (Figure~\ref{Fig:4}E), reveals robustly recurrent behavior (red shows nearby points). Diagonal bands in this recurrence plot occur when the trajectory has parallel segments separated in time\cite{eckmann1987recurrence} and thus reveal the existence of cyclic flux along the trough of the energy landscape. 
	
	Figure~\ref{Fig:4}F shows the time evolution of the distribution of points -- a kymograph -- driven by both stochastic and deterministic processes in the infinite data limit. Using asymmetrical $\mathbf{M}$ we reconstruct the kymograph in Figure~\ref{Fig:4}F, using a simulation of just four revolutions around the trough (Figure~\ref{Fig:4}G, red line). Despite undersampling, the estimated kymograph closely approximates the results obtained in the infinite data limit. This strength of the asymmetric transition matrix is indispensable for the reconstruction of the manifold on the basis of short recordings of neuronal activity.  
	
	The key mathematical insight is that, after appropriate normalization, eigenvalues and eigenvectors of $\mathbf{M}$ approximate those of the discrete version of Eq.~\ref{EMM} \cite{nadler2006diffusion}. Asymmetry of $\mathbf{M}$ admits complex eigenvalues that give rise to cyclic fluxes (second term of Eq.~\ref{EMM}). $\mathbf{M}$ can therefore be directly implemented to simulate Eq.~\ref{EMM}. 
	
	At steady state, all fluxes except for the one with the largest associated complex eigenvalue dampen out and may be safely ignored. Thus, eigenmode decomposition of $\mathbf{M}$ naturally leads to dimensionality reduction that preserves the most salient dynamics. The fundamental advantage of asymmetrical DMs is that the relationship  between the observed states of the system (elements of $\mathbf{M}$) and the relevant macroscopic dynamical variable -- phase -- is given by the  eigenvectors of $\mathbf{M}$. 
	
	In the case of the  system in Figure~\ref{Fig:4} phase is the only relevant variable, while in \textit{C. elegans} (Figure~\ref{Fig:3}) one additionally needs to specify the which loop the system is currently in (flux ID). 
	
		\begin{figure}[H]
			\centering
			\includegraphics[width=\textwidth]{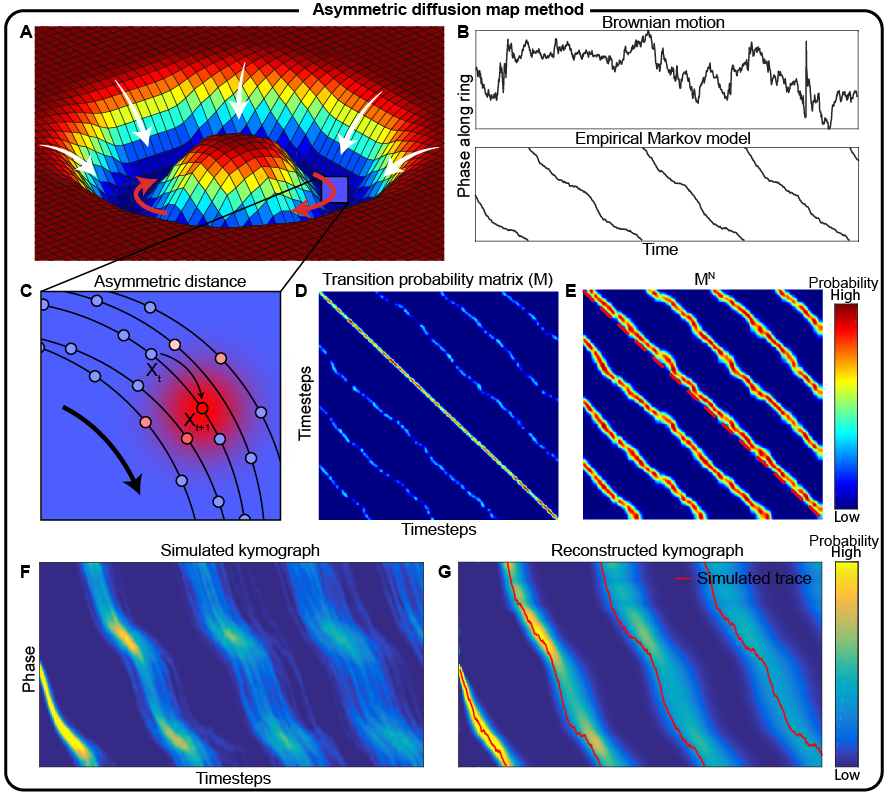}
			\phantomcaption
		\end{figure}
		\begin{figure}[H]
			\ContinuedFloat
			\caption{\textbf{Asymmetric diffusion maps recover phase of the cyclic flux.}
				\small{\textbf{A)} A simulated system consists of a potential (mesh) with associated Brownian motion (white arrows), and cyclic flux (red arrows). \textbf{B)}  Simulation of the system without cyclic flux has no persistent phase velocity (top). With cyclic flux the direction of the rotation is consistent but, because the magnitude of the flux is not uniform, the velocity is variable (bottom). \textbf{C)} Schematic of asymmetric transition probability. The arrow indicates the direction of preferred flux. Transition probabilities from the state of the system at time $t$ ($\mathbf{X}_t$) are computed as in Eq.~ \ref{DiffusionMap} (red $\rightarrow$ high probability; blue $\rightarrow$ low probability). Transition to the next observed state of the system $\mathbf{X}_{t+1}$ is most likely but transitions to parallel trajectories are also possible. \textbf{D)} Asymmetrical transition probability matrix $M$ computed for all pairs of states. $M$ is sparse because for each point, transition probability is nonzero just for $k=12$ nearest neighbors.  The parallel diagonal bands highlight the high level of recurrence in the system. \textbf{E)} Exponentiation of $M$ further highlights the asymmetry. The dashed line shows the main diagonal illustrating that the parallel bands are off center. \textbf{F)} Simulated time evolution of distribution of points (kymograph) created by averaging and smoothing 1000 individual simulations of the system. \textbf{G)} Reconstruction of the kymograph using $M$ estimated from a single simulation of four revolutions around the trough (red line).}
				\label{Fig:4}}
		\end{figure}
		
		\clearpage
	
	\section*{Relevant macroscopic variables are revealed by delay embedding observed neuronal activity.}

	It is typically assumed that there is one-to-one correspondence between a snapshot of neuronal activity and  position in phase space of the brain\cite{schneidman2006weak}. Yet, at any given moment, neuronal activity is influenced by myriad unobserved biophysical variables. Thus, trajectories originating from identical instantaneous neuronal activity patterns can  evolve in different directions depending on the state of these unobserved variables. Indeed, trajectories in the space spanned by neuronal activity are tangled (Figures \ref{Fig:2}G, \ref{Fig:allPca}, \ref{Fig:allPcaDiff}). 
	
	While it is impossible to simultaneously observe all of the microscopic variables that define even the simplest nervous system, one can nonetheless  reconstruct  system dynamics from the observed neuronal activity based on Takens’ delay embedding theorem\cite{takens1981detecting,packard1980geometry,eckmann1987recurrence,sugihara2012detecting} (Figure~\ref{Fig:5}C). 
	
	To illustrate how delay embedding works we use a simple system (Supplement S5)  of two coupled differential equations (Figure~\ref{Fig:5}A) summarized in a single function $f$ 
	\begin{equation} \label{fullSystem}
	\frac{d\mathbf{X}}{dt} = f(O,H).
	\end{equation}  
	The position $\mathbf{X}$ in phase space is given by the pair of dynamical variables ($O$, $H$) and only $O$ is experimentally observed (Eq.~\ref{toyO} and \ref{toyH}). However, given a suitably chosen delay time $\tau$ and the number of embeddings $n$, one can construct a function $\phi$ only on the basis of $O$ such that the dynamics given by $f$ and $\phi$ are essentially equivalent\cite{takens1981detecting}.
	\begin{equation} \label{delayEmbedding}
	\frac{d\mathbf{X}}{dt} = \phi( O(t), O(t - \tau), O(t - 2\tau),...,O(t - n\tau)),
	\end{equation}  
	For the system given by Eq.~\ref{fullSystem} this approach is shown in \ref{Fig:5}A-D.
	
	Delay embedding is a key aspect in the manifold construction for the \textit{C. elegans} nervous system. The eigenvectors of $\mathbf{M}$ map delay embedded neuronal activity onto the state space spanned by phase and flux ID. Without delay embedding, $\mathbf{M}$ fails to produce meaningful predictions (Figure~\ref{Fig:5}E). Thus, there does not appear to be a simple one-to-one mapping between snapshots of neuronal activity and states of the brain. 
	
	
	This conclusion does not rest only on the manifold construction method (Supplement S3). In Figure~\ref{Fig:5}F-I we attempt to directly predict the onset of a backing bout (Figure~\ref{Fig:1} vertical dashed red line) given an antecedent snapshot of neuronal activity. This prediction works well within a single worm with either all recorded neurons (Figure~\ref{Fig:5}F) or the 15 neurons identified in all worms (Figure~\ref{Fig:5}G). Thus, within each worm instantaneous neuronal activity and behavior are closely related. Yet, the prediction degrades dramatically across worms (Figure~\ref{Fig:5}H-I blue) suggesting that the relationship between neuronal activity and behavior is inconsistent among individuals. This is consistent with failure of EMM and diffusion maps constructed on the basis of raw neuronal activity and argues strongly that these failures are not due to dimensionality reduction or binning. 
	
	Remarkably, delay embedding restores the ability to predict across worms (Figure~\ref{Fig:5}F-I orange).  Thus, data in Figure~\ref{Fig:5}E-I strongly imply  that the instantaneous neuronal activity varies among individuals and is not sufficient to specify the state of the brain. Yet, the slow dynamics extracted by the manifold are universal.
	
	To illustrate this further,  we plot normalized activation of representative neurons as a function of behavioral phase in different worms. This reveals that even within a clonal population, individual worms exhibit qualitative differences between activation of some neurons (Figure~\ref{Fig:5}J, Figures \ref{Fig:badNeurons}) while other neurons (Figure~\ref{Fig:goodNeurons}) are consistent  among worms. Furthermore, the same identified neuron can be consistent among worms in one behavior but not in others (Figure \ref{Fig:pValues}). This makes it unlikely that the observed differences in neuronal activations among worms can be attributed to an artifact of imaging  or neuron identification. 
	
		\begin{figure}[H]
			\centering
			\includegraphics[width=\textwidth]{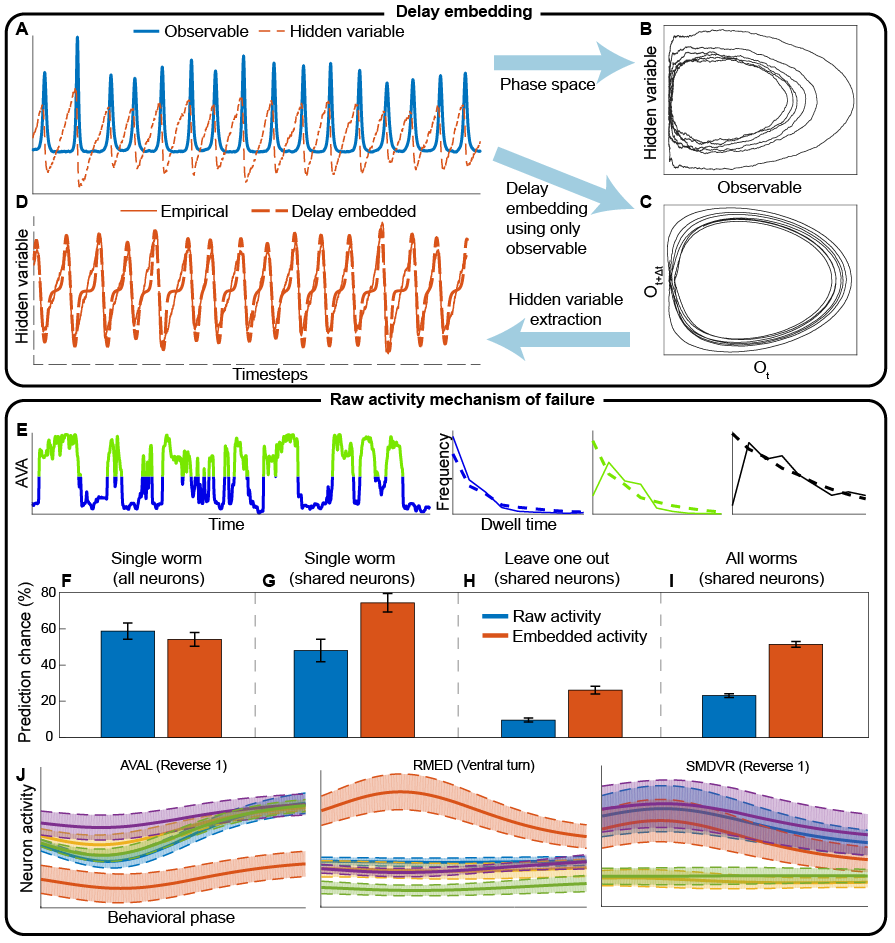}
			\phantomcaption
		\end{figure}
		\begin{figure}[H]
			\ContinuedFloat
			\caption{\textbf{Necessity of Delay Embedding for Extraction of Universal Neuronal Dynamics.}
				\small{\textbf{A)} Time evolution of the system given by Eq.~\ref{fullSystem} is described by two variables ($O$ $\rightarrow$ blue; $H$ $\rightarrow$ orange). Only $O$ is experimentally observed \textbf{B)} Phase space plot of the system in \textbf{A}. \textbf{C)} Phase space reconstructed using approach in Eq~\ref{delayEmbedding}. For visualization purposes only the dimensionality is reduced by projecting onto the first two PCs. \textbf{D)} The hidden variable $H$ is reconstructed from the delay-embedded manifold. \textbf{E)} Manifold constructed as in \ref{Fig:3} on the basis of the 15 shared neurons across worms, fails to qualitatively (left) or quantitatively (right) predict behavioral statistics (forward blue; backward green; backing bouts black) without delay embedding. \textbf{F-I)} Comparison of predictive power on initiations of backing bouts (red dashed line, Figure~\ref{Fig:1}) using different combinations of worms and delay embedding. Leave one out analysis calculates the prediction model based on the data of 4 worms and predicts events for the excluded worm. Blue bars are computed using raw activity while orange bars use delay-embedded data. p-values using two-sample $t$-test: single worm (all neurons) $\rightarrow$ 0.85, single worm (shared neurons) $\rightarrow$ $10^{-5}$, leave one out (shared neurons) $\rightarrow$ $10^{-38}$ and all worms (shared neurons) $\rightarrow$ $10^{-84}$ \textbf{J)} Average normalized traces of specific neuronal activations during specific behaviors.Dotted lines show 95\% confidence intervals. Traces are colored according to the identify of the worm.To account for differences in duration of individual behaviors, time is converted to phase which runs from the initiation to termination of each behavior. While in each invidual worm activation is consistent, there are qualitative consistent differences between worms with respect to certain neurons. p-values of traces coming from the same distribution (1-way ANOVA): AVAL $\rightarrow$ $2.1\times 10^{-7}$, RMED $\rightarrow$ $3.7\times10^{-17}$, SMDVR $\rightarrow$ $1.0\times10^{-9}$}
				\label{Fig:5}}
		\end{figure}
		
	\clearpage
	
	\section*{Discussion.}
	
	Here we developed a method for extracting salient dynamical features from complex, multivariate, nonlinear, and noisy time series which arise in a variety of experimental settings such as gene expression, signal transduction, financial fluctuations, and others. We apply this method to whole brain imaging in \textit{C. elegans} to demonstrate its use in simulating activity of the nervous system and predicting in a model-free fashion switches between different behaviors across individuals. The manifold in \textit{C. elegans} nervous system is composed of two cyclic fluxes corresponding to forward and backward locomotion. While the system is in either one of the fluxes, its fate is entirely predictable. Yet, in the neighborhood where fluxes merge, the behavior cannot be clearly predicted and stochastic forces play a stronger role. This implies that the region where the two cyclic fluxes merge is a decision point where the nervous system is most susceptible to noise and/or sensory inputs. 
	
	The manifold shape is conserved among individuals. Yet, this universality is not observed at the microscopic level of individual neurons. While some neurons exhibit consistent activation patterns across different worms, many others do not. This striking observation is not without precedent. Hodgkin-Huxley models of conductances measured in individual AB neurons in crustacean stomatogastric ganglion exhibit bursting akin to the biological neuron. However, averaging these measurements across different AB neurons yields models that fail to burst\cite{golowasch2002failure}. 
	This suggests that differences between individual AB neurons\cite{goldman2001global,prinz2004similar} or individual worms are not simply random deviations from a common template that can be averaged away at the microscopic level. This is because from the evolutionary perspective the relevant variables are macroscopic. 
	
	While hereditary variability between individuals is indeed implemented at the microscopic level, selection operates at the macroscopic level of organismal behavior\cite{lassig2008biological} embodied by the global dynamics of the brain. Thus, there is no explicit selective pressure for each worm to produce  identical neuronal activation during behavior. While undoubtedly there are important constraints imposed by the biomechanics of the animal, the connectome, and other variables, any degenerate solution that gives rise to the appropriate macroscopic dynamics is equally valid from an evolutionary standpoint. Even in the simplest nervous systems, it is not possible to appropriately experimentally constrain a detailed biophysical model of the brain. Yet, asymmetrical diffusion maps can be used to extract the relevant macroscopic variables that allow for predictions of brain activity valid across individuals.

	\bibliography{MainText}

	
	\section*{Acknowledgments}
	We thank Sarah Friedensen, Guillermo Cecchi, Marcelo Magnasco, Drew Hudson, Tom Joseph, Manuel Zimmer, and Max Kelz for critically reading the manuscript. We also thank Manuel Zimmer and his lab for sharing their recordings of neuronal activity.

	\renewcommand{\thefigure}{S\arabic{figure}}
	\renewcommand{\theequation}{S\arabic{equation}}
	
	\section*{Supplementary}
	
	\section*{S1. Neuronal Imaging}
	
	\textbf{Calcium imaging. }
	
\begin{figure}
		\centering
		\includegraphics[width=\textwidth]{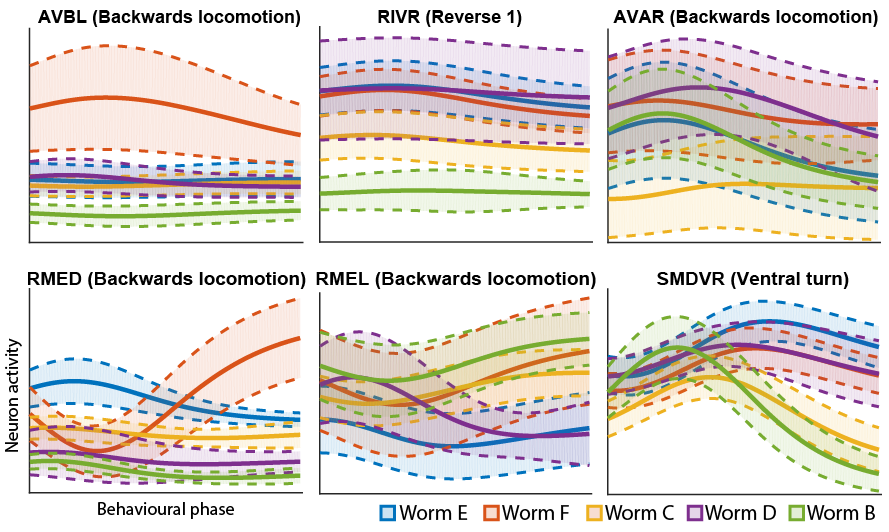}
		\caption{\textbf{Representative neurons which are different across individuals}
			Average traces of specific neurons during specific behaviors in all 5 worms demonstrate individual variance in the clonal population. Each color trace is a different worm. Dotted lines show 95\% confidence intervals. p-values of traces coming from the same distribution (1-way ANOVA): AVBL $\rightarrow$ $0.0056$, RIVR $\rightarrow$ $0.0018$, AVAR $\rightarrow$ $0.0027$, RMED $\rightarrow$ $3.2\times10^{-22}$, RMEL $\rightarrow$ $1.6\times10^{-8}$, SMDVR $\rightarrow$ $7.4\times10^{-10}$
			\label{Fig:badNeurons}}
	\end{figure}

\begin{figure}
		\centering
		\includegraphics[width=\textwidth]{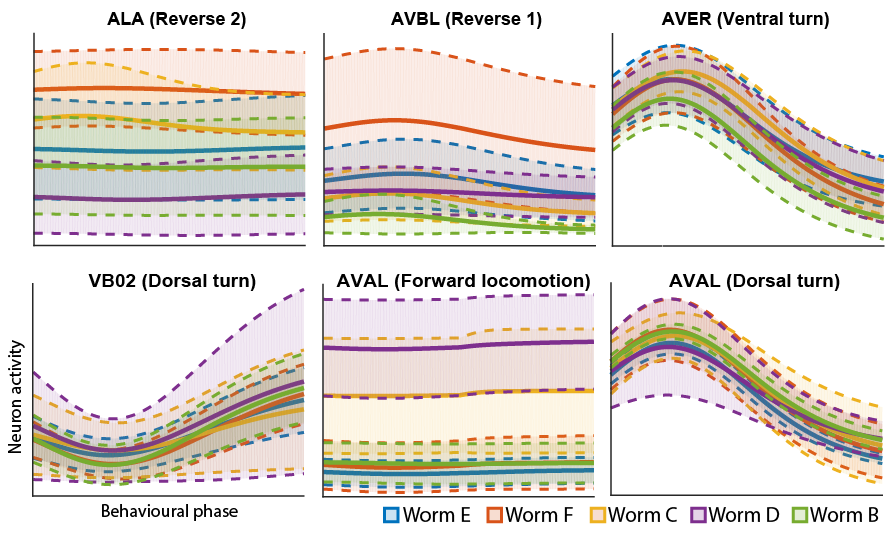}
		\caption{\textbf{Representative neurons which are similar across individuals}
			Average traces of specific neurons during specific behaviors in all 5 worms show that some neurons do have invariant stereotyped dynamics. Each color trace is a different worm. Dotted lines show 95\% confidence intervals. p-values of traces coming from the same distribution (1-way ANOVA): ALA $\rightarrow$ $0.63$, AVBL $\rightarrow$ $0.77$, AVER $\rightarrow$ $0.38$, VB02 $\rightarrow$ $0.42$, AVAL (Forward locomotion) $\rightarrow$ $0.48$, AVAL (Dorsal turn) $\rightarrow$ $0.70$
			\label{Fig:goodNeurons}}
	\end{figure}

\begin{figure}
		\centering
		\includegraphics[width=\textwidth]{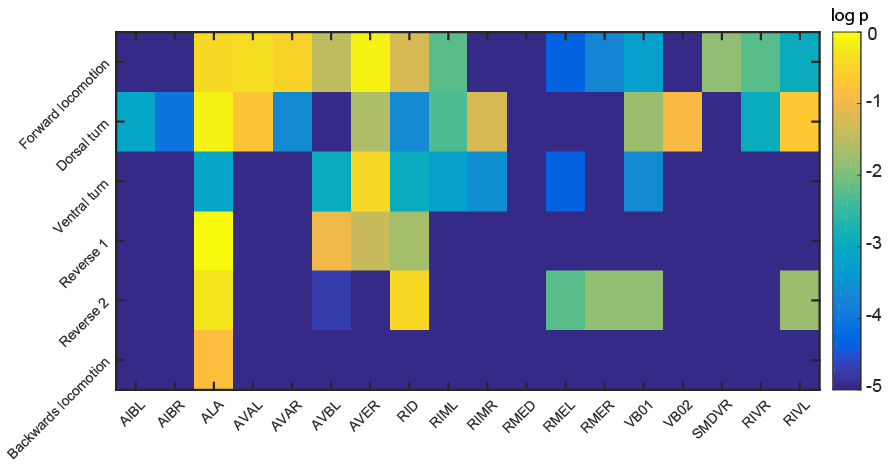}
		\caption{\textbf{p-values for all neuron/behavior combinations}
			Each column is a specific neuron over the 6 identified behaviors. p-values are calculated as in Figures~\ref{Fig:badNeurons} and \ref{Fig:goodNeurons}. Note that some neurons tend to be highly conserved across all behaviors (ALA, RID), while most are only highly similar in a few behaviors. RMED varies consistently in all behaviors.
			\label{Fig:pValues}}
	\end{figure}

	Our analyses use whole-brain single-cell-resolution Ca2+ imaging data published by Kato et al., 2015. The deviation of fluorescence from baseline ($\Delta F/F$) is considered as a proxy for neuronal activity. Each of the five worms were immobilized in a microfluidic device\cite{schrodel2013brain} under environmentally constant conditions. The 107 to 131 neurons detected in each worm span all head ganglia, all head motor neurons and most of the sensory neurons and interneurons along with most of the anterior ventral cord motor neurons\cite{white1986structure}. Of the identified neurons for each worm there is a subset of 15 neurons which were successfully identified across all five worms which are used when generalizing models across multiple worms. The 15 unambiguously identified neurons are: AIBL, AIBR, ALA, AVAL, AVAR, AVBL, AVER, RID, RIML, RIMR, RMED, RMEL, RMER, VB01, VB02. This set of neurons is used in all cross worm analyses. There are 3 more neurons that are very likely (though not unambiguously) identified in all 5 worms -- SMDVR, RIVR and RIVL. These neurons are considered as candidates for representative neurons for Figure~\ref{Fig:5}J and Figures \ref{Fig:badNeurons} and \ref{Fig:goodNeurons}.
	We adopt the same behavior states defined by Kato et al., 2015. The four primary behavioral states are forward locomotion, turns (or FALL), RISE and backwards locomotion (Figure~\ref{Fig:1}). More details of the experiment can be found in Kato et al., 2015.
	
	\textbf{Preprocessing.}
	
	No preprocessing is applied to the data for the Brownian motion and EMM model formulations (Figure~\ref{Fig:2}). For the Fokker-Planck diffusion map method the data are smoothed with a Gaussian filter $\sigma = 1$. Time derivatives are calculated using the $\texttt{diff}$ function. Further, the data $\Delta F/F$ is z-scored in each channel using the $\texttt{zscore}$ function of MATLAB. These values were chosen to maximize the results of each method, but we find that the results are robust to the details of this preprocessing procedure.
	
	Cross worm analyses pose two unique concerns. First, there are only 15 neurons that were identified for in all five animals. These 15 neurons account for the majority of the variance in the data and serve as our cross worm basis (Figure~\ref{Fig:pcaProjections}). Second, the five time series data sets are disjoint. The switches from one set to another are padded with a set of 50 zero values. This length was chosen so that the delay embedding of one worm would never interfere with the embedding from another worm.
	
	Autocorrelation of each channel was computed separately. We are concerned with the autocorrelation of all channels simultaneously and so use the minimum autocorrelation across all channels, which was found to be $\tau \approx 10$ frames. 
	
	\section*{S2. Stochastic Dynamical Systems (Fokker-Planck) }
	
	\textbf{Background.}
	
	Because neuronal systems are constantly exposed to internal and external noise, the most sensible approach is to model the system probabilistically\cite{yan2013nonequilibrium},
	\begin{equation} \label{stochastic}
	\frac{d \mathbf{X}}{dt} = \mathbf{F}(\mathbf{X}) + \mathbf{\epsilon},        
	\end{equation}
	where $\mathbf{F}(\mathbf{X})$ is the driving force, $\mathbf{X}$ is the position in state space and $\mathbf{\epsilon}$ is noise. Since we are concerned with the overall statistics of the system we will consider not the trace of a single trajectory, but the flow of probability about the neuronal space\cite{pathria1996statistical}. The law of probability conservation,
	\begin{equation} \label{conservation}
	\frac{dP(\mathbf{X})}{dt}  = -\nabla \mathbf{J} (\mathbf{X},t),       
	\end{equation}
	states that the change in probability $P$ is due to the local flux, $\mathbf{J} (\mathbf{X},t)$, in that region. In systems with homogeneous (constant in space) noise the probability flux is defined by the equation
	\begin{equation} \label{probability}
	\mathbf{J}(\mathbf{X},t) = \mathbf{F}(\mathbf{X})P - \nabla P ,     
	\end{equation}
	where $\mathbf{F}(\mathbf{X})$ is the driving force of the system at state $\mathbf{X}$.
	Our first model, Brownian motion, assumes that the only dynamics acting on the system are diffusion.  The steady state distribution of this model much match the true probability distribution observed in the data. At steady state the probability distribution no longer changes with time and so,
	\begin{equation} \label{steadystate}
	\frac{dP(\mathbf{X})}{dt} = 0 = \nabla \mathbf{J} (\mathbf{X},t),   
	\end{equation}
	where $P$ is the steady state probability distribution function of the model. In the purely diffusive model we assume that flux of the system vanishes at all points, $ \mathbf{J} (\mathbf{X},t) = 0$, which satisfies the constraint that the divergence of the flux be 0. The relationship between driving force and flux further implies that
	\begin{equation} \label{brownian}
	\mathbf{F}(\mathbf{X})=D\frac{\nabla P(\mathbf{X})}{P(\mathbf{X})}. 
	\end{equation}
	The above uses the trivial solution to Eq.~\ref{steadystate} ($\mathbf{J}=0$), but another class of solutions exist when the flux does not vanish at steady state\cite{yan2013nonequilibrium}. These solutions have flux that is purely cyclic,
	\begin{equation} \label{cyclic}
	\mathbf{J}(\mathbf{X},t) = \nabla \times  A,
	\end{equation}
	where $A$ is an arbitrary vector field. Such fluxes are divergence free, and will always form complete loops. In this case the driving force is,
	\begin{equation} \label{fokkerplanck}
	\mathbf{F}(\mathbf{X})=\mathbf{D}\frac{\nabla P(\mathbf{X})}{P(\mathbf{X})} + \frac{\mathbf{J}(\mathbf{X})}{P(\mathbf{X})},
	\end{equation}
	Where $\mathbf{J}(\mathbf{X})$ is the flux at steady state. Now the driving force is made of two distinct terms. The first term corresponds to diffusion, while the second corresponds to a deterministic cyclic flux. It is in this way that both stochastic and deterministic elements enter into our models.
	
	\textbf{Brownian motion model.}
	
	The Brownian motion model is defined by a right stochastic matrix having the property that each row in the matrix sums to one. This kind of matrix is also known as a Markov matrix and can be interpreted as a transition probability matrix from each neuronal state, $i$, to all other neuronal states, $j$. We limit the diffusion of the system to adjacent pairs of states (bins in a plane spanned by PC1 and PC2), and so the probability of transitioning from state $i$ to state $j$ is proportional to the driving force at that state. This gives us a formula for each off diagonal element of the transition matrix,
	\begin{equation} \label{browniantransition}
	\mathbf{T}(i,j) \sim \frac{\left | p_i - p_j\right | }{p_i},
	\end{equation}
	where $\mathbf{T}(i,j)$ are the elements of the transition matrix and $p_i$ is the probability of a given state in the observed probability distribution. The equation for the transition probability is the discrete equivalent of the driving force (Eq.~\ref{brownian}). Finally, the matrix is scaled by some value so that all rows have a sum less than one and the diagonals are filled to bring the row sum to one. This scaling step represents the level of noise in the system and has no influence on the long term statistics of the model or our results.
	Eq.~\ref{brownian} ensures that the Markov matrix for the Brownian motion model satisfies the detailed balance condition, $\pi_i P(X_{t+1} = j | X_{t} = i) = \pi_j P(X_{t+1} = i | X_{t} = j)$, where $P$ is the probability of transitioning from one state to another. This means that the Markov chain produced by this matrix will be reversible, $P(X_t = i | X_{t-1} = j) = P(X_{t+1} = i | X_t = j)$, which is equivalent to the purely diffusive ($\mathbf{J} = 0$) case of solutions to Eq.~\ref{steadystate}.
	
	\textbf{Empirical Markov Model.}
	
	In the EMM the transition probability matrix is defined by the observed transitions in the data,
	\begin{equation} \label{transition}
	T_{ij}=\frac{\left \| s_i \rightarrow s_j \right \|}{\left \| s_i \right \|},
	\end{equation}
	Where $\left \| s_i \rightarrow s_j\right \|$ is the number of times the system transitions from state $i$ to state $j$ and $\left \| s_i \right \|$ is the total number of times the system is found in state $i$. 
	The resulting Markov chain can be irreversible. This is because the data are not required to have detailed balance, and so the Markov matrix is able to have complex eigenvalues. In a Markov matrix the eigenvalues can be thought of as the modes along which the system decays to its steady state distribution. This is because the a Markov matrix is the time evolution operator of the system given by the differential equation,
	\begin{equation} \label{timeevolution}
	\Delta\mathbf{X}_t = \mathbf{X}_t\mathbf{L}, 
	\end{equation}
	where $\mathbf{L}$ is the Markov matrix. The solution to this differential equation is,
	\begin{equation} \label{solution}
	\mathbf{X}_t = \mathbf{X}_oe^{\mathbf{L}t},
	\end{equation}
	where $\mathbf{X}_o$ are the initial condition of the system. Alternatively this equation can be rewritten in terms of the eigenmodes of $L$,
	\begin{equation} \label{eigenevolution}
	\mathbf{X}_t = \sum_{i}^{ } c_i e^{\lambda_it}\phi _i,
	\end{equation}
	where $\lambda_i$ are the eigenvalues, $\phi_i$ are the eigenvectors and $c_i$ are the initial conditions of the system projected onto the $i$-th eigenvector. 
	$\mathbf{L}$, under a broad range of conditions, has a single eigenvalue with $\lambda = 1$. This corresponds to an assertion that such systems always come to a single steady state. The associated eigenvector corresponds to the steady state distribution of the system. When eigenmodes are complex Eq.~\ref{eigenevolution} becomes an equation of a decaying wave in the plane spanned by a pair of complex conjugate eigenvectors. As discussed in the main text, these decaying spirals correspond to the cyclic flux of Eq.~\ref{fokkerplanck}. In the long time limit all eigenmodes with eigenvalues much less than 1 damp out, but complex modes with eigenvalues near 1 heavily shape the dynamics of the system even in the long time limit. These eigenmodes are used to identify the cyclic fluxes of neuronal activity. 
	
	\section*{S3. Model performance measures }
	
	\textbf{Dwell Time Statistics.}
	
	Dwell time statistics are calculated by setting a threshold on the simulated AVA neuron. The threshold is chosen by inspection, and any time point above the threshold is assumed to belong to backing locomotion, while any time point below the threshold belongs to forward locomotion. Backing bout times are periods in which the animal stops backing up and immediately backs up again. These events are defined as periods in which the forward locomotion state fails to last for more than 30 frames (~10 seconds). Dwell time distributions are shown in Figure~\ref{Fig:1}-\ref{Fig:3}, \ref{Fig:5}. Statistics for single worm models using EMM are shown in Figure~\ref{Fig:EMMStats}. 
	
	\textbf{Manifold-Free behavioral prediction.}
	
	We selected a neuronal event that is likely to be highly correlated with behavior, and thus informative of the relevant neuronal dynamics\cite{fetz1992movement,michaels2016neural,georgopoulos1986neuronal,salinas2001correlated}. behaviorally, this event is the onset of a backing bout. 
	An average activation pattern was constructed by aligning the time series of neuronal activity to each initiation of  a backing bout and averaging across them (akin to spike triggered average). In the simplest case, this average activation pattern is a single snapshot of neuronal activity immediately preceding the initiation of the backing bout. Delay embedding (see below and main text) involves taking several snapshots of neuronal activity separated by a fixed time interval. Average activation patterns of multiple lengths (number of snapshots) were evaluated for their ability to predict (see below) the initiation of the bout. The snapshots were separated by $\tau \approx 10$ (~3 seconds). This value was  chosen based on the autocorrelation of neuronal activity. 
	
	To evaluate the ability of the average activation pattern to predict the onset of backing bout, the  average activation pattern was convolved with neuronal activity and smoothed. The output of this convolution (score) was then used to calculate the prediction probability shown in Figure~\ref{Fig:5}F-I (see below). 
	
	In order to predict the onset of the behavioral event an appropriate threshold value of the score has to be chosen. From an information-theoretic perspective, an optimal threshold is a threshold that gives the highest separation between the distribution of scores associated with the initiation of backing bout and those that are not. 
	
	To construct these distributions only local maxima of the score were considered. This minimizes the spurious effects of noise and compensates for the fact that the behavioral events of interest are very rare in the data. To construct the distribution of scores associated with the behavioral event, we found peaks closest to the behavioral event within  $\left | t_{peak}-t_{event} \right |\leq 10\approx 3s$. The amplitude of these peaks is used to construct the distribution of true events $\theta =1$.  All other peaks are considered to be false events, $\theta =0$. 
	
	The probability of correctly identifying a behavioral event given a specific $X_{thres}$ is 
	\begin{equation} \label{threshold}
	p(\theta=1\mid X\geq X_{thres})=\frac{\left \| X_{\theta=1}\geq   X_{thres} \right \|}{\left \| X_{\theta=1} \right \| + \left \| X_{\theta=0} \geq X_{thres} \right \|},
	\end{equation}
	where $X_{\theta=1}$ are the scores of true events, $X_{\theta=0}$ are scores of false events and $X_{thres}$ is the score cutoff threshold, and $\left \| \cdot \right \|$ denotes the number of elements in the set. Optimal threshold is found as argmax of Eq.~\ref{threshold} with respect to$X_{thres}$. 
	
	The average activation pattern is constructed using 50\% of the events in the data. This training set is used to find  $X_{thres}$ and the remaining 50\% are used to determine the accuracy of the prediction based on the scores exceeding $X_{thres}$. To obtained standard deviation of the prediction probability, the training set is bootstrapped.
	
	\textbf{Time to transition analysis.}
	
	Phase along the cyclic flux in the manifold is binned. Points within each bin are identified. Each point is characterized by two independent values: time since the onset of the behavior and bin ID. The null hypothesis here is the expected time to behavioral transition based solely the dwell time distribution. This corresponds to finding the survival function given by the right tail of the integral of the dwell time distribution (to infinity).
	\begin{equation} \label{nullprobability}
	P_{null}(t) = \sum_{i}^{N} \frac{P(t + t_i)}{N}.
	\end{equation} 
	Where $t_i$ is the time since onset of the behavior, and  $P(t)$ is the probability of the transition occurring at time $t$. For the manifold-based prediction the distribution of times until behavioral switch is explicitly found in the data. This analysis applied to backward locomotion is shown in Figure~\ref{Fig:3} and for forward locomotion in Figure~\ref{Fig:forwardPhase} and \ref{Fig:forwardEntropy}. 
	
	\section*{S4. Manifold Finding}
	
	\textbf{Delay embedding and true phase space.}
	
\begin{figure}
		\centering
		\includegraphics[width=\textwidth]{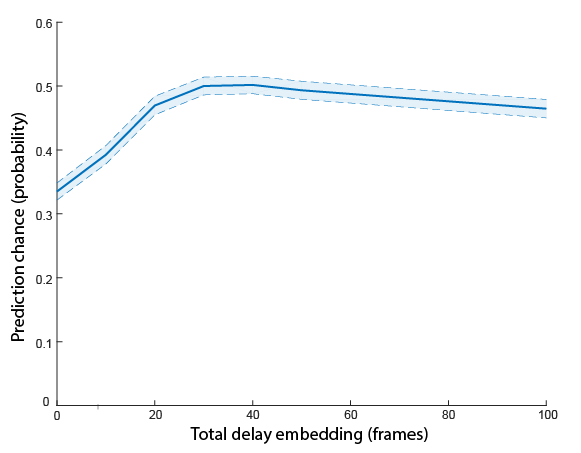}
		\caption{\textbf{Benefits of delay embedding are not infinite}
			Plot of predictive power (using manifold-free behavioral prediction) as in Figure~\ref{Fig:5}. Varying the amount of delay embedding increases the predictive power to a point (~40 frames). After this point, the added dimensionality begins to have adverse effects and the predictive power begins to fall.
			\label{Fig:maxDelay}}
	\end{figure}

	In the case of \textit{C. elegans} benefits of delay embedding are not infinite -- \textit{i.e}. more data does not always yield better predictions (Figure~\ref{Fig:maxDelay}). This conclusion is independent of the manifold construction method. This observation suggests the existence of a true phase space, where trajectories are maximally coherent, and minimal trajectory overlap occurs. 
	
	While several algorithms for finding a good delay embedding parameters exist\cite{packard1980geometry}, they have significant limitations for highly noisy and short (3000 to 4000 frames per animal). To improve our chances of finding a reasonable embedding, rather than embedding just the raw neuronal activity, we embedded the adjoint space formed by the raw neuronal activity and its derivative (akin to position and velocity). While there are some differences depending on the specifics of the parameters of the delay embedding, the results are fairly robust to changes in the parameters. For the figures in the main manuscript we used:
	\begin{itemize}
		\item[] $\Delta \tau = 10$ (Number of frames to delay in each delay embedding)
		\item[] Number of $\tau= 5$  (Total number of delay embeddings)
	\end{itemize}
	For the toy model in Figure~\ref{Fig:4} we used:
	\begin{itemize}
		\item[] $\Delta \tau = 1$ (Number of frames to delay in each delay embedding)
		\item[] Number of $\tau = 1$ (Total number of delay embeddings)
	\end{itemize}
	For the toy model in Figure~\ref{Fig:5}A we used:
	\begin{itemize}
		\item[] $\Delta \tau = 149$ (Number of frames to delay in each delay embedding)
		\item[] Number of $\tau = 20$ (Total number of delay embeddings)
	\end{itemize}
	\textbf{Diffusion mapping. }
	
	Diffusion maps refer to a class of methods in which distances between points (in state space) are cast as transition probabilities\cite{nadler2006diffusion,coifman2006diffusion} -- nearby points have high transition probability. The transition probability between nearby points can be approximated by a Gaussian diffusion kernel,
	\begin{equation} \label{dmkernel}
	k_\varepsilon (\mathbf{X},\mathbf{Y})=exp\left (-\frac{\left \| \mathbf{X} - \mathbf{Y} \right \|_2^2}{4\varepsilon }  \right ),
	\end{equation} 
	where $\varepsilon$ is a parameter of the algorithm, and $\left \| \cdot  \right \|_2$ is the standard euclidean norm. 
	
	It turns out that, diffusion maps converge to the normalized graph Laplacian, the Fokker-Planck\cite{nadler2006diffusion} operator or the Laplace-Beltrami operator depending on the normalization of $k_\varepsilon$,
	\begin{equation} \label{normalizedkernel}
	k_\varepsilon^\alpha  (\mathbf{X},\mathbf{Y})=\frac{k_\varepsilon  (\mathbf{X},\mathbf{Y})}{q_\varepsilon^\alpha (\mathbf{X})q_\varepsilon^\alpha (\mathbf{Y})},
	\end{equation} 
	where $q_\varepsilon(\mathbf{X})$ is the probability of point $\mathbf{X}$ at steady state, and $k_\varepsilon$ is the pre-normalized kernel. The Fokker-Planck operator corresponds to the case where $\alpha =\frac{1}{2}$. This normalization is used for the rest of this discussion.
	
	While diffusion mapping is an extremely powerful tool it is not directly applicable to neuronal activity because the temporal component of the time series data is not utilized in the analysis\cite{lian2015multivariate}. As a consequence, the diffusion kernel and resulting transition matrix is symmetric. Therefore, we will seek to use the temporal component of the data to form an asymmetric transition probability matrix.
	
	We start by redefining the diffusion kernel in order to make use of the time series data. We accomplish this simply by centering the kernel on the next empirically observed data point as follows 
	\begin{equation} \label{ourkernel}
	k_{FP}  (\mathbf{X}_t,\mathbf{Y})=exp\left (-\frac{\left \| \mathbf{X}_{t+1}-\mathbf{Y} \right \|_2^2}{2 \sigma^2}  \right ).
	\end{equation} 
	Where $\mathbf{X}_t$ is the position of the system in state space at time $t$, $\mathbf{Y}$ is a nearby position in state space, and $\sigma$ is a normalization term (see below). 
	
	There are two other differences in our kernel compared to the original version\cite{coifman2006diffusion}. First, the normalization term, $\sigma$, is defined by using the amount of local noise at the two points $\mathbf{Y}$ and $\mathbf{X}_{t+1}$. The local noise, $\sigma_l(x)$ is approximated by taking the standard deviation of the velocity of the trace in some small time window around the point of interest. This means that $\sigma^2$ is now the product of two standard deviations, which is a natural symmetric scaling factor for the diffusion-based kernel. The method is robust to the exact value of time window used, and smoothing techniques can also be employed as the noise level is assumed to be a function of position, and the time series data moves smoothly through space. Second, in order to remove the dependence of $k_\varepsilon$ on the $\varepsilon$ parameter we include in $\sigma$ the mean value from all of the local neighbors found throughout the full data set. Again, this normalization is quite robust and only serves to automatically pick a $\varepsilon$ that is known to be in a reasonable range. Putting these together yields the full kernel,
	\begin{equation} \label{fullkernel}
	k_{FP}  (\mathbf{X}_t,\mathbf{Y}_t)=exp\left (-\frac{\left \| \mathbf{X}_{t+1}-\mathbf{Y}_t \right \|_2^2}{2 \sigma_l(\mathbf{X}_{t+1})\sigma_l(\mathbf{Y}_t)\left \langle k_{FP} \right \rangle _{\mathbf{X}\mathbf{Y}}}  \right  ),
	\end{equation} 
	where $\left \langle k_{FP} \right \rangle _{\mathbf{X}\mathbf{Y}} $ is the mean value of $k_{FP}$ over all data points and associated nearest neighbors.
	
	The time series data is assumed to have coherent trajectories, however, in the limit of finite data with progressively finer time steps the above kernel will fail to appropriately model diffusion between coherent trajectories. This is because there will be many points from the same trajectory (due to the small time steps) that are very close to $\mathbf{X}_{t+1}$ while points from separate yet coherent trajectories may be relatively far away. In order to overcome this problem, the notion of nearest neighbors is adjusted to instead be nearest trajectories. The kernel for $\mathbf{X}_t$ to all other points is calculated and then the best point is added to the nearest trajectories list (this will always be point $\mathbf{X}_{t+1}$ since it has a kernel value of 1). Any points that are within some minimum number of time steps of $\mathbf{X}_{t+1}$ are removed from consideration. This assures that the next nearest point will come from a trajectory other than the original trajectory. Then the closest remaining point is added to the nearest trajectories list and the process repeats for $k$ nearest trajectories. Both this minimum time frame and the number of nearest trajectories to be considered are important parameters of the method and can heavily influence the results. The minimum time frame must be chosen to be long enough that neighboring trajectories are indeed independent trajectories and so a good estimate can be found using the autocorrelation or the medium recurrence time of the time series data. The number of nearest trajectories is a trade off between needing enough data to be able to cluster the coherent trajectory bundles appropriately and not forcing the method to group trajectories that are not in fact in the same coherent bundle. However, both of these issues are resolved in the large data limit.
	
	\noindent For the reconstruction of worm manifold we used:
	\begin{itemize}
		\item[] Local trace size = 6 (Number of points to approximate local noise)
		\item[] Number of neighbors = 12 ($k$ nearest trajectories to link in the transition probability matrix)
		\item[] Minimum time frame = 50 (Minimum number of frames to wait for the nearest trajectories)
	\end{itemize}
	For the toy model in Figure~\ref{Fig:4} we used:
	\begin{itemize}
		\item[]Local trace size = 4 (Number of points to approximate local noise)
		\item[]Number of neighbors = 10 ($k$ nearest trajectories to link in the transition probability matrix)
		\item[]Minimum time frame = 50 (Minimum number of frames to wait for the nearest trajectories)
	\end{itemize}
	
	\textbf{Phase Identification.}
	
	The key objective of the method is to identify a small set of parameters by which to characterize the state of the brain across individual worms. In terms of the Fokker-Planck system given by the diffusion map, the key parameter is phase. Phase is given by the eigenvector associated with the largest complex eigenvalue as all other eigenmodes do not strongly influence the dynamics in the long term. This phase can be directly computed from the transition probability matrix formed by exponentiating the matrix constructed as in Eq.~\ref{fullkernel}. 
	
	If there are multiple cyclic fluxes as in \textit{C. elegans} CNS, then in addition to the phase one needs to also know the identify of the flux. This latter problem is addressed in clustering section (see below).
	
	\textbf{Trajectory clustering.}
	
	Any standard clustering algorithm will suffice, and this section will only detail one of many possible choices\cite{rubinov2010complex} that can be used. We did not explore the effects of the choice of clustering and suspect that, as is the case with many clustering applications, the best choice will depend on the specifics of the dataset. We use a maximum modularity algorithm\cite{newman2006modularity} on the transition probability matrix defined in Eq.~\ref{fullkernel}. By construction, the transition probability matrix is sparse (only transitions in local neighborhoods are considered). Therefore, in its raw form the system given by this matrix will not explore the manifold sufficiently as it will be trapped in each individual isolated neighborhood. To overcome this problem the matrix is exponentiated N times until a minimum fraction of elements of each row are non-zero (25\% in the worm data). Conceptually this corresponds to finding the evolution of the system after $N$ time steps. Specific choice of $N$ does not have a strong influence on the results, so long as the resultant matrix is not too sparse. 
	
\begin{figure}
		\centering
		\includegraphics[width=\textwidth]{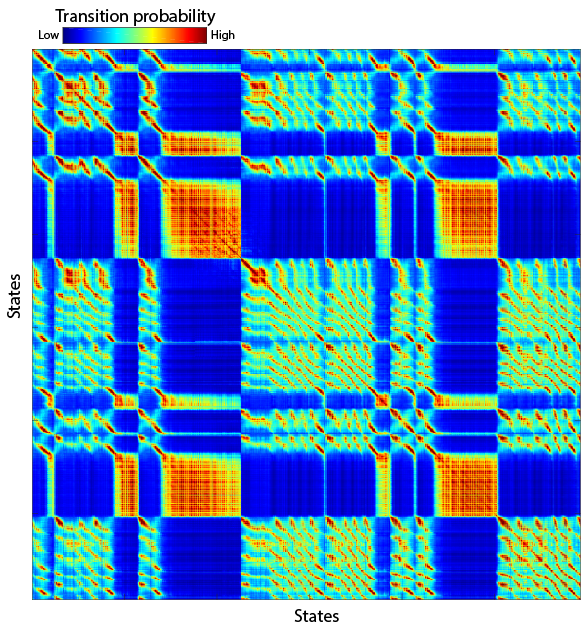}
		\caption{\textbf{Asymmetric diffusion  matrix modified for clustering of trajectories}
			Asymmetric diffusion matrix constructed using diffusion kernel as in Eq.~\ref{fullkernel}. This matrix has been modified for clustering by convolving each row of the matrix with all other rows and taking the maximum resulting value. In this way, the diagonal bands that make up trajectories tend to cluster together more tightly.
			\label{Fig:diffusionMatrix}}
	\end{figure}

	Two major features are found in the transition probability matrix (Figure~\ref{Fig:diffusionMatrix}): patches and diagonals. Square patches identify locations where the system exhibits Brownian motion near a point attractor. Diagonal traces identify coherent trajectories.
	
	The square patches are already suitable for community detection. If two elements of the matrix belong to the same point attractor, they will be found in the same square patch. The situation is slightly more complex for coherent trajectories identified by diagonal bands of high transition probability. To determine whether two elements of state space belong to the same coherent trajectory, we compute the maximum correlation of each row (distances from each element of state space) and time lagged copies all the other rows $max(corr(row_i, shift(row_j,t)))$. Where $row_i$ is the $i$th row, shift moves all elements in the row $t$ steps to the right and the maximum is taken over all $t$. This newly formed matrix has the same dimensions as the original transition probability matrix. We apply standard maximum modularity clustering using the $community\_louvain$ function from the Brain Connectivity Toolbox to this matrix\cite{rubinov2010complex}.
	
	\textbf{Manifold simulation.}
	
\begin{figure}
		\centering
		\includegraphics[width=\textwidth]{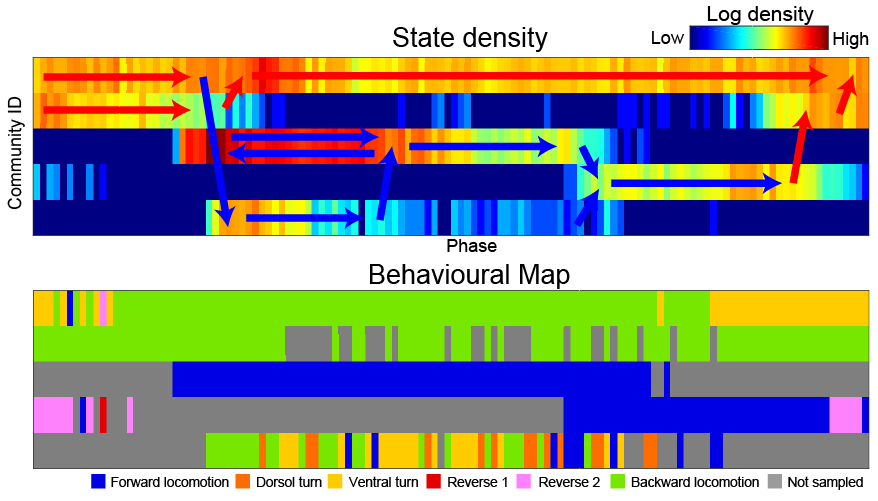}
		\caption{\textbf{Manifold space density and behavioral localization}
			The matrix of the kind shown in Figure~\ref{Fig:diffusionMatrix} was subjected to clustering. This identified 5 clusters (communities) each of which is shown by a row in the plot. X-axis is the phase computed from the asymmetric diffusion matrix constructed according to Eq.~\ref{fullkernel}. Top plot shows density of points. Red arrows show the flux of the system corresponding to a backwards locomotive behavior, and blue arrows show the flux of the system corresponding to a forward locomotive behavior. Bottom plot shows behavioral states (colored as in the main text).
			\label{Fig:manifoldSpace}}
	\end{figure}

	Our ability to construct EMM (Figure~\ref{Fig:2}) in the PCA space hinges on the relative simplicity of the \textit{C. elegans} CNS allowing for significant dimensionality reduction. In high dimensional spaces, direct binning is not possible as each bin would be too sparsely occupied to give meaningful information concerning the probability distribution or the transition probabilities between neighboring bins. Yet, as we show in Figure~\ref{Fig:2} the PCA space is not universal across animals and is thus limiting in terms of our ability to simulate brain dynamics. In contrast, the phase and cluster identity of each point defines a natural basis on which to build a low dimensional representation of the time series data that is densely populated. In fact, our simulations of brain dynamics are essentially EMM model in the proper space of flux ID and phase. The density of points in this space can be seen in the plot of log probability density function over the parameters of phase and community identity (Figure~\ref{Fig:manifoldSpace} top). The space also retains information about the behavioral state of the animal, as shown by the alignment of behavior to each community identity (Figure~\ref{Fig:manifoldSpace} bottom).
	
	The simulated AVA traces in Fig \ref{Fig:3}C are created by simulating the system with the same procedure as the EMM using this parameterized space. A map from parameterized state space, $(ID, \phi) \in \mathbf{\Theta}$, to the true neuronal phase space, $\mathbf{X}$, is defined by averaging the neuronal space position of all points found in each parameter space bin,
	\begin{equation} \label{manifoldspace}
	\mathbf{\Theta } \rightarrow  \mathbf{X} \:(ID,\phi )_i\mapsto\left \langle \mathbf{X}_t \in (ID,\phi )_i  \right \rangle.
	\end{equation} 
	
	\textbf{Manifold reconstruction. }
	
	In order to reconstruct the shape of the manifold in neuronal space (Figure~\ref{Fig:3}) we need some way in which to determine the mean trajectory followed by each trajectory bundle. We can find this mean by using a combination of the two parameters discussed in the previous sections. The phase provides a continuous parameter that runs along the flow of the trajectory bundle, and the cluster identity provides an insurance that the trajectory bundles with overlapping phases are treated separately.
	
	In order to ensure that our phase parameter continuously covers all space, we construct a Gaussian smoothed phase density histogram in order to find the bounds of that trajectory by including only those phases that have density larger than some cutoff. The method is robust to the exact value used.
	
	These averaged trajectory start and end points are then used to segment the time series data and align it by the phase over this particular trajectory. The mean trajectory is then found by averaging the neuronal position of points using weights from a sliding Gaussian window over the phase values. The method is robust to the exact sigma used.
	
	The resulting mean trajectories form disjoint bundles in the phase space. In order to construct the smooth manifolds as shown in Figure~\ref{Fig:3} the bundles are joined together by interpolating a spline (over both position and direction) from the end of one bundle to the beginning of the next bundle. The qualitative accuracy of the interpolation is checked vs. the raw time series data. Nonetheless, these extensions are not necessary for quantitative analyses -- which are all done in $(ID, \phi) \in \mathbf{\Theta}$
	
	\textbf{Average behavioral trajectories.}
	
	To compare the differences in patterns of neuronal activation within a single worm and across multiple worms we synchronize activity time to behavioral outputs and average over this synchronized time. For each point in a particular instance of a behavior we set its synchronized time to be $t_{sync}(t_i) = (t_i - t_{start}) / (t_{end} - t_{start})$. Where $t_i$ is the time of the point, $t_{start}$ is the time of the start of the instance of behavior and $t_{end}$ is the time of the end of the instance of behavior. This time warping insures that $t_{sync}$ ranges from 0 to 1 over all instances of behavior, and thus can be used to average trajectories from multiple instances of the same behavior.
	
	p-values were calculated using only the first PC of the traces. Time series data of each behavioral trace was projected onto the first PC calculated from the concatenated matrix of all traces for each neuron/behavior pair. Each individual trace was mean subtracted before performing PCA. p-values were then computed using 1-way ANOVA.
	
	\section*{S5. Toy Systems}
	
	\textbf{Delay Embedding.}
	
	The system used to highlight delay embedding is given by the two-parameter ($O$ and $H$) system,
	\begin{equation} \label{toyO}
	O' = H\cdot O + \epsilon,
	\end{equation} 
	\begin{equation} \label{toyH}
	H' = a(b-O^{2}) + \epsilon,
	\end{equation} 
	where $a$, $b$ and noise $\epsilon$ are selected to keep the system near trajectory bundles. We use $a = 0.01$, $b = 1.2$ and noise is Gaussian with mean zero and $\sigma = 0.1$. These parameters were chosen such that the system takes roughly 5000 timesteps to recur. The two equations above completely define the phase space of the system, along with future and past trajectories (up to noise) (Figure~\ref{Fig:5}).
	
	\textbf{Model Stochastic Differential Equation. }
	
	The system used in Figure~\ref{Fig:4} was created by constructing an energy landscape and flux term and then simulating using the stochastic differential equation,
	\begin{equation} \label{modelsystem}
	\Delta \mathbf{X} = \lambda_U \nabla U(\mathbf{X}) + \lambda_JJ(\mathbf{X},t)+\epsilon,
	\end{equation} 
	where $\lambda_U$ and $\lambda_J$ are strength coefficients for the diffusion and flux terms respectively, and $\epsilon$ is noise. The energy landscape is defined by single ring formed by tracing a Gaussian around the circle and taking the maximum value at each increment. The flux follows the equation,
	\begin{equation} \label{modelflux}
	\mathbf{F}(\mathbf{X})\sim\widehat{\theta }\frac{1+\alpha sin(\theta )}{( r-\mu_r  )^{\beta }},
	\end{equation} 
	where $\widehat{\theta}$ indicates flux in the polar direction, and $\mu_r$ is the radial coordinate with the highest flux. $\mu_r$is set to be equal to the radius of the trough formed by the energy landscape, and the coefficients  $\lambda_U$, $\lambda_J$, $\alpha$ and $\beta$ are set such that the system makes ~1 complete cycle every 1000 frames.
	
	\clearpage
	
	\begin{figure}
		\centering
		\includegraphics[width=\textwidth]{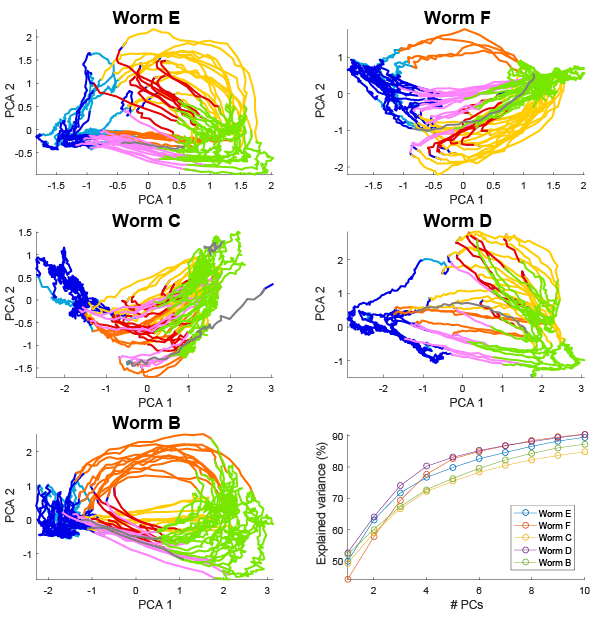}
		\phantomcaption
	\end{figure}
	\begin{figure}
		\ContinuedFloat
		\caption{\textbf{PCA projections of pan-neuronal imaging data}
			Projection of neuronal activity onto the first 2 PCs of each individual worm. All neurons recorded in each worm are included in the analysis. Data are colored by behavioral states as in the main text. behaviors localize into mostly coherent patterns. Note, however, that the shape of the data differs significantly between worms. Furthermore, the PCs themselves are different between worms. Plot of the explained variance for the first 10 PCs for each worm (bottom right). Note that the first 2 PCs explain approximately 60\% of the variance in each  worm. 
			\label{Fig:pcaProjections}}
	\end{figure}
	
	\begin{figure}
		\centering
		\includegraphics[width=\textwidth]{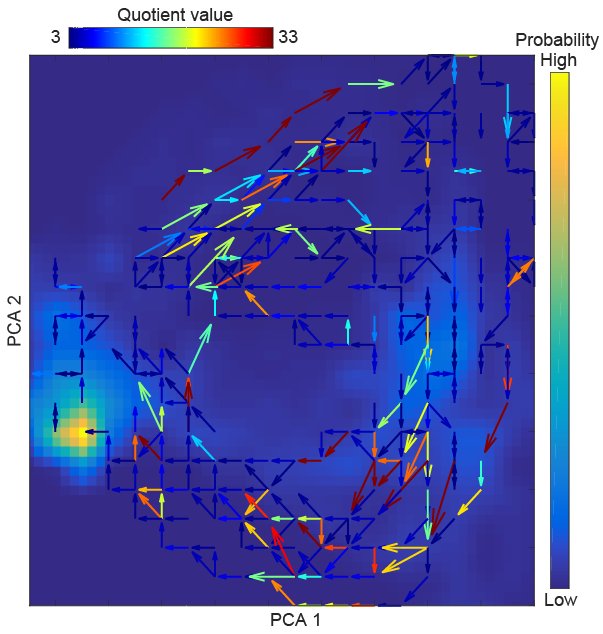}
		\phantomcaption
	\end{figure}
	\begin{figure}
		\ContinuedFloat
		\caption{\textbf{Comparison of Brownian motion and EMM}
			Element-wise quotient of the transition matrices (EMM / Brownian motion) is computed. If the value at position $(i,j)$ of the resultant matrix exceeds threshold, an arrow is drawn from point $i$ to point $j$. The arrow is colored from blue (EMM $\approx $ Brownian motion) to red (EMM $>>$ Brownian motion). Large values of the quotient are mostly concentrated along the transitions between the two energy minima. This flux is, on average, confined to a clockwise cycle. 
			\label{Fig:quotient}}
	\end{figure}
	
	\begin{figure}
		\centering
		\includegraphics[width=\textwidth]{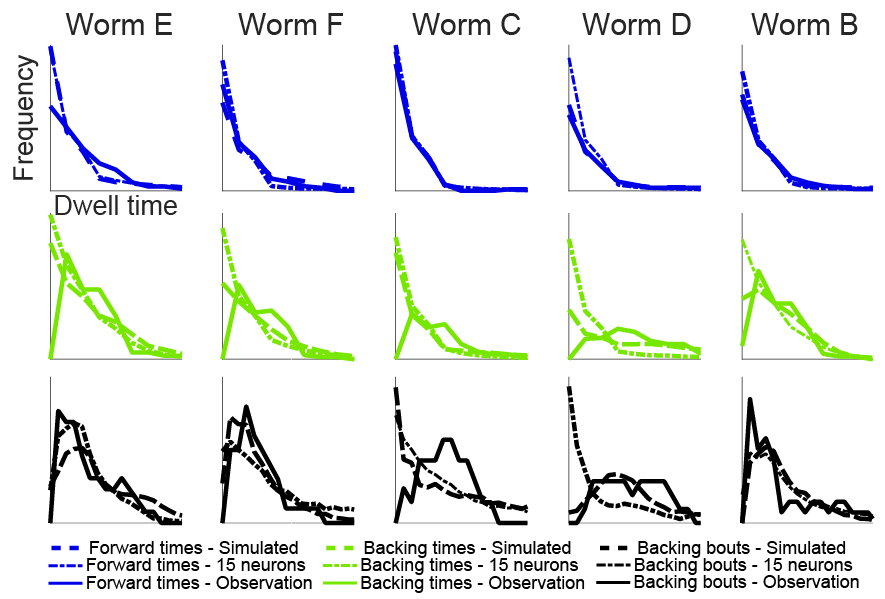}
		\phantomcaption
	\end{figure}
	\begin{figure}
		\ContinuedFloat
		\caption{\textbf{Dwell time statistics predicted by EMM constructed for each  worm individually}
			Dwell time statistics of forward locomotion (blue), backwards locomotion (green) and backing bouts (black) observed experimentally (solid line) and predicted by EMM (dashed lines) in each  worm. EMM was constructed as in the main manuscript on the basis of the first two PCs of neuronal activity. Either the full neural set (dashed) or the common 15 neuron subset (dot dash) were used. In each case, PCs were calculated individually for each worm. The main manuscript shows data from Worm B (all neurons). This worm is most closely predicted by EMM and makes the best possible case for this approach. Forward locomotion is well approximated in all worms. Backward locomotion is generally less well approximated by EMM and appears more stochastic than is experimentally observed. Backing bouts are well approximated in all worms with the exception of worm C. Which is likely due to the high level of tangling of trajectories found in its projection onto PC1 and PC2 (Figure~\ref{Fig:pcaProjections}).
			\label{Fig:EMMStats}}
	\end{figure}
	
	\begin{figure}
		\centering
		\includegraphics[width=\textwidth]{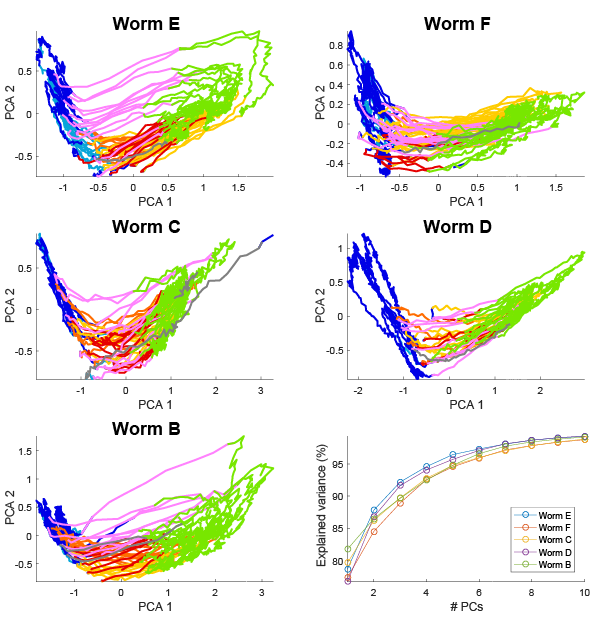}
		\phantomcaption
	\end{figure}
	\begin{figure}
		\ContinuedFloat
		\caption{\textbf{Projections of 15 shared neurons on the first two PCs computed individually for each worm}
			Projection of the shared set of 15 neurons from each individual worm onto the first 2 PCs calculated from its data. Contrast this with Figure~\ref{Fig:pcaProjections} which shows the same analysis for all neurons observed in each worm. Data are colored by the same behavioral states as assigned in the main text. Note that as in Figure~\ref{Fig:pcaProjections} the PCs are computed individually for each worm and thus will not generally be the same between worms.  Plot of the explained variance for the first 10 PCs (bottom right). Note that the first 2 PCs explain approximately 85\% of the variance for each worm. 
			\label{Fig:pca15Projections}}
	\end{figure}
	
	\begin{figure}
		\centering
		\includegraphics[width=\textwidth]{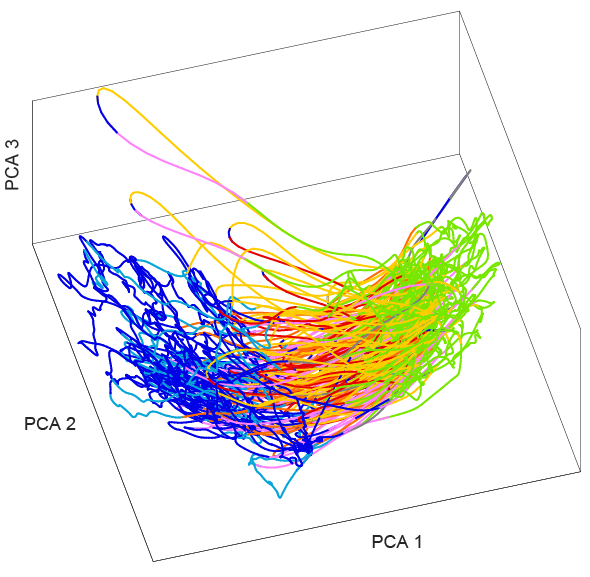}
		\phantomcaption
	\end{figure}
	\begin{figure}
		\ContinuedFloat
		\caption{\textbf{15 shared neuronal activity of all worms projected onto same PCA basis}
			Activity of each of the 15 neurons identified in all worms was transformed to z-scores. Thus normalized data from each worm was concatenated and subjected to PCA. Projection onto first 3PCs is shown. Data are colored by the same behavioral states as assigned in the main text. When data from all worms is plotted on the same axes, trajectories form a tangle, and behavioral states are not clearly separated. This tangling of the trajectories explains why EMM is not directly generalizable across worms even if the number of dimensions is increased from 2 to 3. 
			\label{Fig:allPca}}
	\end{figure}
	
	\begin{figure}
		\centering
		\includegraphics[width=\textwidth]{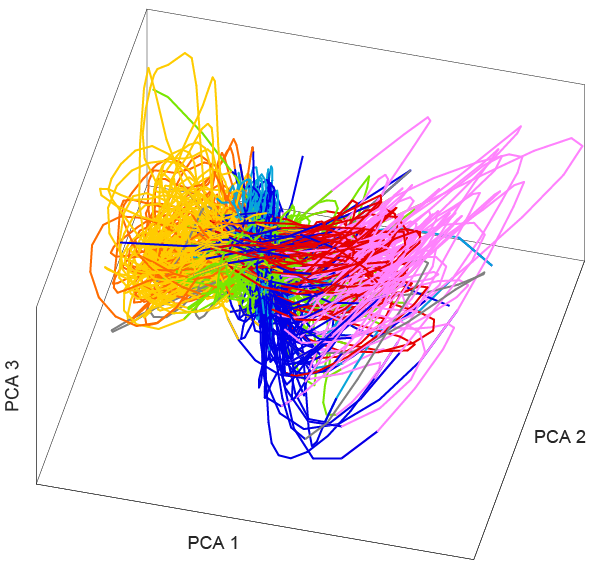}
		\phantomcaption
	\end{figure}
	\begin{figure}
		\ContinuedFloat
		\caption{\textbf{Derivative of 15 shared neuronal activity of all worms projected onto same PCA basis}
			It was noted by Kato et al\cite{kato2015global} that derivatives of neuronal activity  separated behavioral states better than raw activity. While this is true, projecting the appropriately normalized derivatives onto a common set of PCs still gives rise to significant trajectory tangling. Note that consistent with Kato et al\cite{kato2015global} the trajectories are separated better than in Figure~\ref{Fig:allPca}. This is because projecting the derivative of neuronal activity in discrete time is closely related to the idea of delay embedding as the calculation of the first derivative requires information about the present and future state of the system. 
			\label{Fig:allPcaDiff}}
	\end{figure}
	
	\begin{figure}
		\centering
		\includegraphics[width=\textwidth]{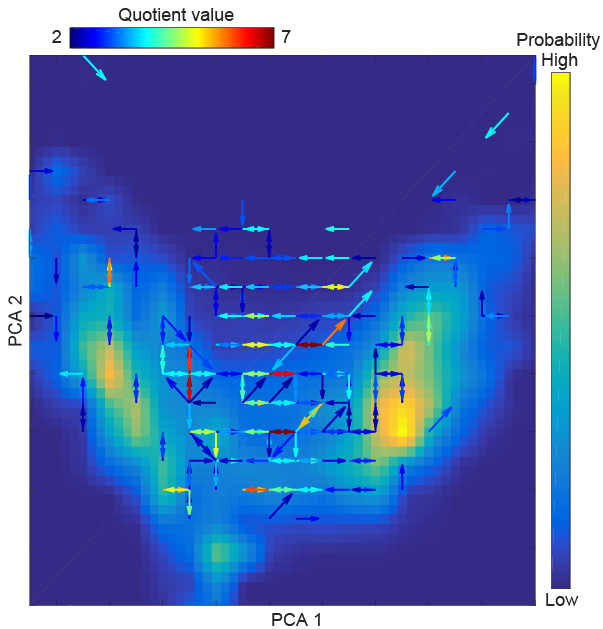}
		\phantomcaption
	\end{figure}
	\begin{figure}
		\ContinuedFloat
		\caption{\textbf{Quotient between Brownian motion and EMM using space spanned by 15 shared neuronal activity}
			Quotient between flux predicted by Brownian motion and experimentally estimated as in EMM for the 15 neurons in all worms projected onto common set of PCs. Unlike Figure~\ref{Fig:quotient} no clear cyclic flux is observed. This is because trajectories in Figure~\ref{Fig:allPca} are tangled. Note that the order of magnitude scale difference for the quotient. This is consistent with the observation that EMM across worms does not perform any better than the stochastic model Figure~\ref{Fig:2}G of the main manuscript 
			\label{Fig:quoitent15}}
	\end{figure}
	
	\begin{figure}
		\centering
		\includegraphics[width=\textwidth]{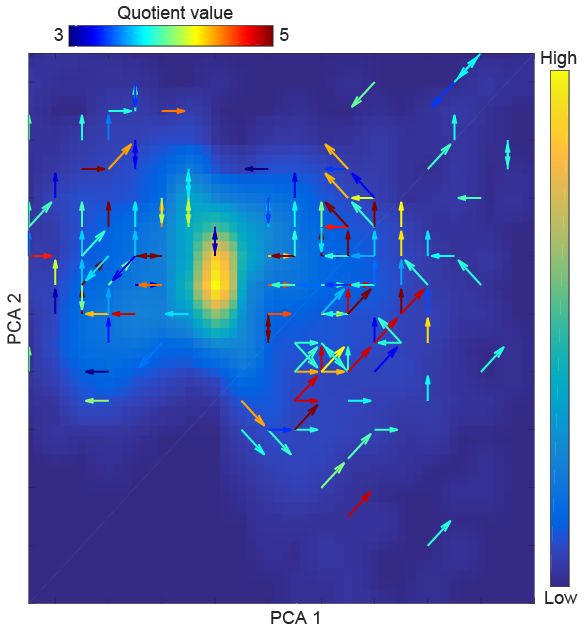}
		\phantomcaption
	\end{figure}
	\begin{figure}
		\ContinuedFloat
		\caption{\textbf{Quotient when using space spanned by derivative of 15 shared neuronal activity}
			Quotient between flux predicted by Brownian motion and experimentally estimated as in EMM for the 15 neurons in all worms projected onto common set of PCs. In this case normalized derivatives of neuronal activity were used as the basis of the PCs. Looking at the derivatives does not restore the cyclic flux clearly present in Figure~\ref{Fig:2}D 
			\label{Fig:quoitent15Diff}}
	\end{figure}
	
	\begin{figure}
		\centering
		\includegraphics[width=\textwidth]{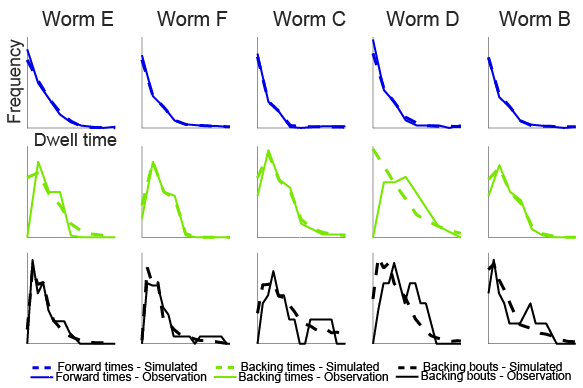}
		\caption{\textbf{Manifold simulation of individual worms}
			Simulations using the manifold found on the basis of all observed neurons in each individual worm. Most worms recapitulate the observed statistics qualitatively, and well approximate the statistics quantitatively. Note that in contrast to EMM (Figure~\ref{Fig:EMMStats}), worm C is well approximated by the manifold. Deviations from the observed dwell times are likely due in part to under sampling (Number of observed behaviors is about 20 for all worms but D, which has about 10 behaviors).
			\label{Fig:mainifoldStats}}
	\end{figure}
	
	\begin{figure}
		\centering
		\includegraphics[width=\textwidth]{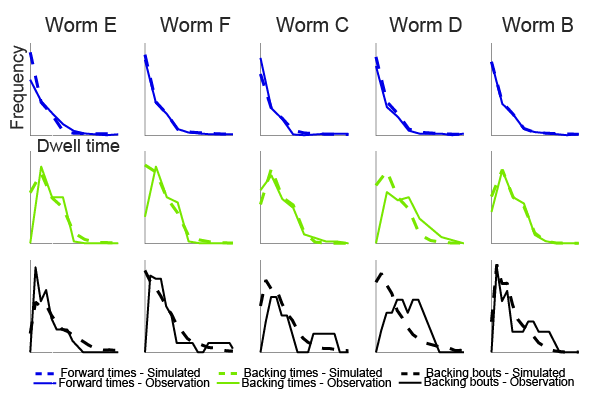}
		\caption{\textbf{Manifold simulation of left out worms}
			Simulations using the manifold found on the basis of 15 shared neurons by leaving each one of the 5 worms in turn. Simulations of neuronal activity given by this manifold are used to predict the behavioral statistics from the left out worm (the data from which was not used in manifold construction). All simulations show structure in their dwell time statistics that qualitatively mimic the statistics found in the left out worm. Worm D is a bit of an outlier in terms of dwell time statistics, which is why the other 4 worms have trouble predicting it. Deviations from the observed dwell times are likely due in part to under sampling (Number of observed behaviors is about 20 for all worms but D, which has about 10 behaviors).
			\label{Fig:leftOutStats}}
	\end{figure}
	
	\begin{figure}
		\centering
		\includegraphics[width=\textwidth]{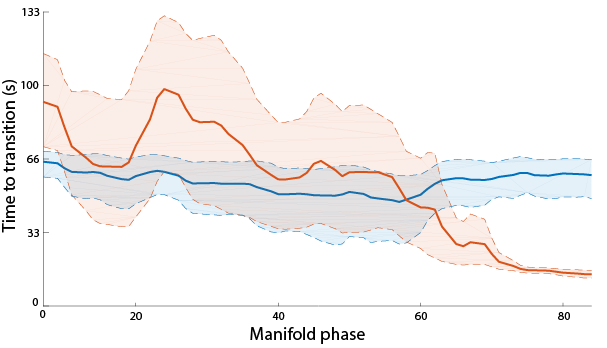}
		\caption{\textbf{Manifold dynamics of forward locomotion}
			Similar analysis to Figure~\ref{Fig:3}F-I of the main text. This subsection of the manifold is taken from the forward locomotion loop and predicts the time at which the worm transitions from forward locomotion to a turn.
			\label{Fig:forwardPhase}}
	\end{figure}
	
	\begin{figure}
		\centering
		\includegraphics[width=\textwidth]{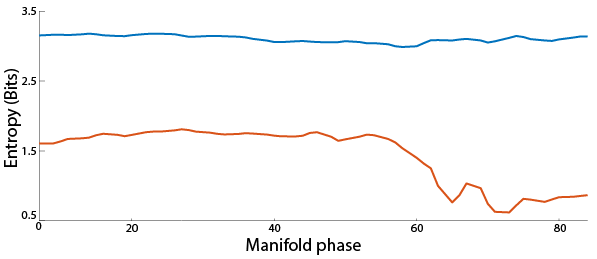}
		\caption{\textbf{Entropy of forward locomotion}
			Similar to the main text the entropy of the manifold based prediction is always lower than the null hypothesis. However, the entropy takes longer to drop off, and does so in a much less linear fashion, due to the high degree of stochasticity found in the forward locomotive trajectory.
			\label{Fig:forwardEntropy}}
	\end{figure}
	
	\begin{figure}
		\centering
		\includegraphics[width=\textwidth]{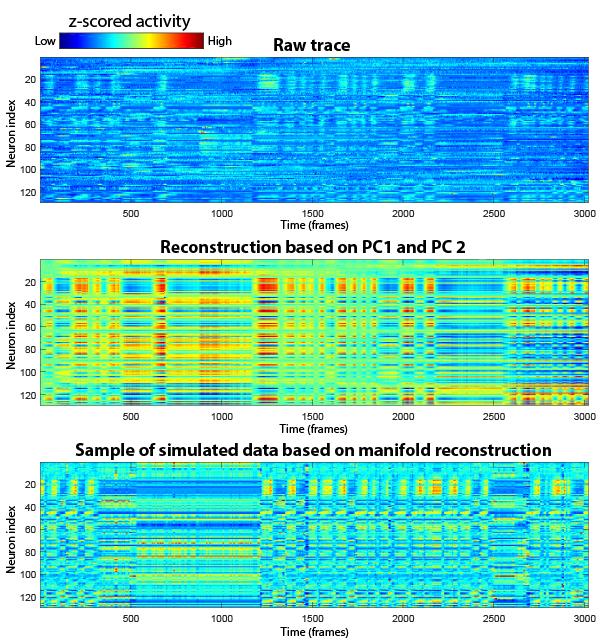}
		\phantomcaption
	\end{figure}
	\begin{figure}
		\ContinuedFloat
		\caption{\textbf{Reconstruction of pan-neuronal dynamics}
			Original data from Figure~\ref{Fig:1}B after z-scoring (top). Reconstructed of the z-scored data based on the projection onto the first two PCs (middle). Note that the main features of the original data are preserved, but the noisier elements are lost. Manifold simulation of the same worm as the above plots using all recorded neurons (bottom). Note that more details of the structure is preserved in this simulation that was preserved in the reconstruction based on the first 2 PCs.
			\label{Fig:reconstruction}}
	\end{figure}
	
	\clearpage
	

\end{document}